\documentclass[MSNbibl,number,citesort,dvips]{arxbj}

\aid{0}
\volume{19}
\issue{4}
\pubyear{2013}
\firstpage{1306}
\lastpage{1326}
\doi{10.3150/12-BEJSP05} 

\makeatletter
\newcommand{\eqref}[1]{(\ref{#1})}
\def\esssup{\mathop{\operatorname{ess\,sup}}}
\makeatother

\begin{document}
\begin{frontmatter}

\title{Probabilistic aspects of finance}
\runtitle{Probabilistic aspects of finance}

\begin{aug}
\author[1]{\fnms{Hans} \snm{F\"ollmer}\thanksref{1}\ead[label=a]{foellmer@math.hu-berlin.de}}%
\and
\author[2]{\fnms{Alexander} \snm{Schied}\corref{}\thanksref{2}\ead[label=b]{schied@uni-mannheim.de}}
\runauthor{H. F\"ollmer and A. Schied} 
\address[1]{Institut f\"ur Mathematik, Humboldt-Universit\"at, 10099 Berlin, Germany.\\ \printead{a}}
\address[2]{Institut f\"ur Mathematik, Universit\"at Mannheim, 68131 Mannheim, Germany.\\ \printead{b}}
\end{aug}


%
\begin{abstract} In the past decades, advanced probabilistic methods
have had significant impact on the field of finance, both in academia
and in the financial industry. Conversely, financial questions have
stimulated new research directions in probability. In this survey
paper, we review some of these developments and point to some areas
that might deserve further investigation. We start by reviewing the
basics of arbitrage pricing theory, with special emphasis on incomplete
markets and on the different roles played by the ``real-world''
probability measure and its equivalent martingale measures. We then
focus on the issue of model ambiguity, also called Knightian uncertainty.
We present two case studies in which it is possible to deal with
Knightian uncertainty in mathematical terms. The first case study
concerns the hedging of derivatives, such as variance swaps, in a
strictly pathwise sense. The second one deals with capital requirements
and preferences specified by convex and coherent risk measures. In the
final two sections we discuss mathematical issues arising from the
dramatic increase of algorithmic trading in modern financial markets.
\end{abstract}

%

\end{frontmatter}

\section{The coin tossing view of finance and the appearance of
Brownian motion}

The systematic use of advanced probabilistic methods in the context of
academic Finance begins in the mid-sixties. It was pioneered at M.I.T.
by Paul Samuelson \cite{Samuelson} and greatly stimulated by the
rediscovery of ``Th\'eorie de la Sp\'eculation'', the doctoral
thesis \cite{Bachelier} of Louis Bachelier, that had been defended in
Paris in 1900 based on a report by Henri Poincar\'e. In this thesis,
Brownian motion makes its appearance as a mathematical model for the
price fluctuations of a liquid financial asset. Arguing that prices
should remain positive, Samuelson proposed to use geometric Brownian
motion, which soon became a standard reference model. In 1973, Black
and Scholes \cite{BlackScholes} and Merton \cite{Merton} derived their seminal formula for the
price of a call-option in this setting.

Why does Brownian motion appear in the financial context? Here is a
first rough argument. At each fixed time, the price of a stock could be
seen as a temporary equilibrium resulting from a large number of
decisions to buy or sell, made in a random and more or less independent
manner: Many coins are thrown successively, and so Brownian motion
should arise as a manifestation of the central limit theorem. This is
the ``Coin-Tossing View of Finance'', as it is called by J.~Cassidy
in \textit{How Markets Fail} \cite{Cassidy}. This rough argument can be
refined by using microeconomic assumptions on the behavior of agents
and on the ways they generate a random demand, and then the application
of an invariance principle typically yields a description of the price
fluctuation as a solution of a stochastic differential equation driven
by Brownian motion or, more generally, by a L\'evy process; see, for
example, \cite{FoellmerSchweizer} and the references therein.

At this point, however, it is instructive to recall the following
caveat of Poincar\'e in \textit{Science et M\'ethode} \cite{P1908} as quoted in \cite{Kirman}:

\begin{quote}When men are in close touch with each other, they no
longer decide randomly and independently of each other, they react to
the others. Multiple causes come into play which trouble them and pull
them from side to side. But there is one thing that these influences
cannot destroy and that is their tendency to behave like Panurge's
sheep. And it is that which is preserved.
\end{quote}
Thus we find, right at the beginning of the use of modern probabilistic
methods in finance, a~warning sign pointing to interaction and herding effects
which may render invalid a straightforward application of the central
limit theorem.

In his ``Three essays on Capital Markets'' \cite{Kreps}, David
Kreps uses a different kind of argument, where geometric
Brownian motion appears as a \textit{rational expectations equilibrium}.
Suppose that agents compute their demand by maximizing expected
utility. If their preferences are given by power utility functions, and
if their subjective expectations are described by geometric Brownian
motion, then the resulting price equilibrium would indeed be a
geometric Brownian motion. Thus geometric Brownian motion
is described as a fixed point for an aggregation problem based on the
preferences and expectations of highly sophisticated agents. Here
again, Poincar\'e's caveat throws some doubt on the assumptions of rationality
implicit in such an argument.

Bachelier himself does not invoke the central limit theorem, nor does
he argue in terms of expected utility. Instead he starts out with a
simple equilibrium argument: ``\textit{It seems that the market, that
is to say, the set of speculators, must not believe in a given instant
in either a rise or a fall, since for each quoted price there are as
many buyers as sellers}''. As a result,
``\textit{the mathematical expectation of the speculator is zero}''.
Stated in modern terms, Bachelier insists that the price process should
be a martingale under a probability measure
$P^*$ which describes the market's aggregate belief. Assuming
continuous paths and adding a stationarity requirement for the
increments, it follows that
the price process is indeed a Brownian motion.

What is the current mainstream view? To begin with, there is a broad
interdisciplinary consensus across departments of Mathematics, Finance,
and Economics that the discounted price fluctuation of a liquid
financial asset should be viewed as a stochastic process $ X =
(X_t)_{0\leq t \leq T}$ on some underlying probability space $(\Omega,
\mathcal{F},P)$. The intuition is typically objectivistic: Such a
probability measure $P$ exists, it can be identified at least partially
by statistical and econometric methods, and it should satisfy certain
\textit{a priori} constraints. These constraints
correspond to some degree of market efficiency. In its strongest form,
market efficiency would require that $X$ is a martingale under $P$. In
the mainstream view, however, a weaker and more flexible version of
market efficiency is assumed,
namely the absence of safe (and not just statistical) arbitrage
opportunities. In other words, the price process should not admit any
trading strategy that
produces a positive expected gain over the risk free return without any
downside risk. If this is made precise in a suitable manner, the
absence of arbitrage opportunities can be characterized by the
existence of an equivalent \textit{martingale measure}, i.e., a
probability measure $P$* equivalent to $P$ such that the properly
discounted price process $X$ is a (local) martingale under $P$*. This
characterization is often called the Fundamental Theorem of Asset
Pricing. A preliminary version appears in Harrison and Kreps \cite
{HarrisonKreps}, and its definitive form is due to Delbaen and
Schachermayer \cite
{DelbaenSchachermayer94,DelbaenSchachermayer98,DelbaenSchachermayer06};
see also Kabanov \cite{Kabanov} and Yan \cite{Yan}.

Thus an economic assumption, namely the absence of arbitrage
opportunities, guarantees that
\[
\mathcal{P}^* \neq\varnothing,
\]
if we denote by $\mathcal{P}^*$ the set of equivalent martingale
measures $P^*$.
Due to well-known results of Jacod, Yor and others in the ``general theory'' of stochastic processes of the 70s and 80s, this
implies that the process $X$ is a semimartingale under the
original measure $P$, and hence a stochastic integrator in the sense of
Bichteler and Dellacherie. This allows one to apply the techniques of
It\^o calculus. Moreover it follows, due to a line of arguments
initiated by Wolfgang Doeblin \cite{Doeblin} and completed by I. Monroe
\cite{Monroe1,Monroe2}, that $X$ is a Brownian motion up to a random
time change. In this way, Brownian motion reappears
in the present general setting, although not necessarily in a very
explicit manner.

\section{Derivatives and the paradigm of perfect hedging}\label{Perfect
hedge Section}

A \textit{derivative}, or a \textit{contingent claim}, specifies a payoff
$H(\omega)$ contingent on the scenario $\omega\in\Omega$ which will
be realized.
For example, a European call option with strike price $c$ and maturity
$T$ has payoff $H(\omega) = (X_T(\omega) -c)^+$. What is the fair price
which should be payed by the buyer of such a contingent claim $H$? In
other words, what is the fair deterministic equivalent to the uncertain
outcome $H$? This is a classical question, and the standard answer goes
back to the founding fathers of probability theory, in particular to
Jacob Bernoulli. It says that you should assign probabilities to the
different scenarios $\omega$ and compute the expected value
\[
E_P[H] = \int H\,dP
\]
of the random variable $H$ with respect to the resulting probability
measure $P$.
Following Daniel Bernoulli \cite{Bernoulli}, one might want to add a
risk premium in order to take account of risk aversion.
More precisely, one could describe risk aversion by a strictly
increasing and concave utility function $u$ and compute the price $\pi
(H)$ of $H$ as the certainty equivalent
$u^{-1}(E_P[u(H)]) $.
The difference $\pi(H)-E_P[u(H)]$, which is positive by Jensen's
inequality, is then interpreted as a risk premium.
But in our present financial context and under the following uniqueness
assumption~\eqref{unique P*}, the basic insight of Black and Scholes
\cite{BlackScholes} and Merton \cite{Merton} leads to a quite different result.
In particular there will be no reason to argue in favor of a risk
premium because the following argument shows that there is no intrinsic risk
in that case.

Consider a financial market model such that $\mathcal{P}^*\neq\varnothing$.
In many situations, and in particular for simple diffusion models such
as geometric Brownian motion, the equivalent martingale measure is in
fact unique, that is,
%
\begin{equation}
\label{unique P*} \bigl| \mathcal{P}^*\bigr| = 1.
\end{equation}
Uniqueness of the equivalent martingale measure implies that the model
is \textit{complete} in the following sense: Any contingent claim $H$ can
be represented, $P$-almost surely, in the form
%
\begin{equation}
\label{replication} H = V_0 + \int_0^T
\xi_t \,d X_t
\end{equation}
with some constant $V_0$ and some predictable process $\xi= (\xi_t)_{0\leq t \leq T}$ such that the stochastic integral makes sense.
For simple diffusion models such as geometric Brownian motion, this
representation follows from It\^o's theorem that functionals of
Brownian motion can be represented as stochastic integrals of Brownian
motion; see \cite{RevuzYor} for the general case. Since the expectation
of the stochastic integral under the equivalent martingale measure
$P^*$ is zero, the constant $V_0$ is given by
$ V_0 = E^*[H]$.

In financial terms, the representation \eqref{replication} amounts to a
perfect replication of the contingent claim by means of a dynamic
trading strategy. Indeed, It\^o's non-anticipative construction of
the stochastic integral allows one to interpret the stochastic integral
in \eqref{replication} as the cumulative net gain generated by the
self-financing trading strategy consisting in holding $\xi_t$ units of
the underlying asset at each time $t$. The constant amount $V_0$ can
now be viewed as the initial capital which is needed for a perfect
replication, or a \textit{perfect hedge}, of the contingent claim. But
this implies that the unique arbitrage-free price of the claim is given by
%
\begin{equation}
\label{unique price} \pi(H)=V_0 = E^*[H],
\end{equation}
since any other price would offer the opportunity for a gain without
any risk. If, for example, the actual price were higher then one could
sell the claim at that price, use the smaller
amount $V_0$ to implement the hedging strategy which generates the
random amount $H$ which has to be paid in the end, and retain the
difference between the price and $V_0$ as a risk-free gain.

Thus, the uniqueness assumption \eqref{unique P*} yields
a simple answer to the problem of pricing and hedging financial derivatives.
Note that the answer only involves the unique equivalent martingale
measure $P^*$. The role of the probability measure $P^*$ is to serve as
a sophisticated consistency check for the pricing of assets, not for
the purpose of prediction. The original probability measure $P$ was
meant to serve that purpose, but here it matters only insofar as it
fixes a class of null sets. As we are going to see in Section \ref
{probability free Section} below, we can actually eliminate $P$
completely if we are ready to restrict the space of possible scenarios.

\section{Incompleteness as a source of new probabilistic problems}

As soon as a financial market model becomes more realistic by admitting
that there are more sources of uncertainty than traded financial
instruments, the equivalent martingale measure is no longer unique, and
this implies
\[
\bigl\vert\mathcal{P}^*\bigl\vert= \infty.
\]
As a result, the paradigm of a perfect hedge breaks down, and intrinsic
risks appear at the level of derivatives.
The model is then called \textit{incomplete}.
From a mathematical point of view, incompleteness has turned out to be
a rich source of new problems in Stochastic Analysis. In particular it
has motivated new versions of probabilistic decomposition theorems such
as the Kunita--Watanabe decomposition and the Doob--Meyer decomposition.

Consider a derivative with non-negative payoff $H$ and maturity date
$T$. An admissible hedging strategy is given by an initial capital
$V_0$ and a predictable process $\xi$ such that the resulting portfolio
process $V$ defined by
%
\begin{equation}
\label{value process} V_t = V_0 + \int
_0^t \xi_s\,dX_s
\end{equation}
remains non-negative. At the maturity date $T$, any such strategy
yields a decomposition
\[
H = V_T + C_T
\]
of the contingent claim into a part which is perfectly hedged, and
hence priced by arbitrage as in the preceding section, and a remaining
\textit{hedging error} $C_T$. Different economic preferences induce
different choices of the strategy, and hence a different decomposition
of the claim.

Suppose one wants to minimize the hedging error in a mean-square sense
with respect to the given probability measure $P$. This will amount to
a projection in the space $L^2(P)$ of the contingent claim $H$ onto a
sub-space of stochastic integrals.
Under the strong form of the efficient market hypothesis, that is $P
\in\mathcal{P}^*$,
this projection problem is solved by using the Kunita--Watanabe
decomposition in the space of square-integrable martingales; see \cite
{FoellmerSondermann}.
If one drops this assumption and considers the case $P \notin\mathcal{P}^*$,
the resulting decomposition problem can often be reduced to an
application of the Kunita--Watanabe representation with respect to a
suitable \textit{minimal} martingale measure; cf. \cite{FoellmerSchweizer0}.
More generally, methods of mean-variance hedging for incomplete
financial markets have been a source of new versions of the
Kunita--Watanabe decomposition and of new results on closure properties
of spaces of stochastic integrals with respect to a semimartingale;
see, for example, the surveys \cite{FoellmerSchweizerEncyclopedia} and
\cite{Schweizer}.

From a financial point of view, however, the mean-variance approach
fails to capture a basic asymmetry: The main purpose is to control the
\textit{shortfall}, defined as the positive part $C_T^+ = (H - V_T)^+$ of
the hedging error. If one insists on keeping the shortfall down to $0,$
then one is led
to a remarkable new extension of the Doob--Meyer decomposition. Consider
a right-continuous version $U$ of the process
%
\begin{equation}
\label{upper U} U_t = \esssup_{P^* \in\mathcal{P}^*} E^*[H\vert
\mathcal{F}_t], \qquad 0 \leq t \leq T.
\end{equation}
Now note that $U$ is a $ \mathcal{P}^*$-supermartingale, that is, a
supermartingale under any $P^* \in\mathcal{P}^*$. As shown in
increasing generality in \cite{ElKarouiQuenez,Kramkov,FoellmerKabanov},
any non-negative $ \mathcal{P}^*$-supermartingale $U$ admits a
decomposition of the form
%
\begin{equation}
\label{optional decomp} U_t = U_0 + \int
_0^t \xi_s\,dX_s -
A_t
\end{equation}
with some increasing optional (but in general not predictable) process
$A$. But the stochastic integral is a $\mathcal{P}^*$-local
martingale, and so \eqref{optional decomp} can be viewed as a new
version of the classical Doob--Meyer decomposition that holds
simultaneously for all $P^* \in\mathcal{P}^*$. In the special case
\eqref{upper U}, this optional decomposition can be interpreted as a
\textit{superhedging} procedure: Starting with the initial capital $V_0 =
U_0$, applying the trading strategy $\xi$ and sequentially withdrawing
the cumulative amount $A_t$ from the generated portfolio value $V_t$
defined in \eqref{value process}, one ends up with the final value $U_T
= H$.
Dually, $U_t$ can be characterized as the minimal capital that is
needed at time $t$ in order to cover the contingent claim $H$ by a
dynamic trading strategy run from time $t$ up to $T$.

The superhedging approach may tie down a large capital amount in order
to stay on the safe side,
and therefore it is usually seen as too conservative. But the
mathematics of superhedging remains important even if zero tolerance
for a shortfall is relaxed. Suppose, for example, that one imposes
some bound for the expected loss $E_P[\ell(C^+_T)]$, defined in terms
of some convex loss function $\ell$. Then the resulting problem of {\it
efficient hedging} can be split into a statistical decision problem,
which is solved by a randomized test $\varphi$, and a dynamic
superhedging problem for the modified claim $\tilde H=\varphi H$; see
\cite{FoellmerLeukert}.

More generally, the efficient hedging problem can be embedded into a
problem of dynamic portfolio optimization
for incomplete financial markets, where the criterion is usually
formulated in terms of expected utility.
There is a rich literature on such dynamic optimization problems, from
the point of view of both optimal stochastic control as in \cite
{KaratzasLehoczkyShreve,KaratzasShreveMF,StoikovZariphopoulou} and
convex duality as in \cite{KramkovSchachermayer1,KramkovSchachermayer2}

Note that in these optimization problems for incomplete financial
markets the probability measure $P$ does come in explicitly, in
contrast to the superhedging approach. But it does so at the level of
preferences, namely in the form of expected utility. As soon as one
admits model uncertainty and considers robust preferences as described
in Section~\ref{risk measure section} below, new problems of robust
optimization arise; see, for example, \cite
{SchiedMOR,SchiedFS,HernandezSchied,F2006} and the survey \cite
{FoellmerSchiedWeber}. Another new direction consists in analyzing the
temporal dynamics of preference structures as in \cite
{MusielaZariphopoulou1,MusielaZariphopoulou2}.

\section{$P$ versus $P^*$}

As we have seen, the standard setting in mathematical finance is
probabilistic, and it involves two types of probability measures.
On the one hand, it assumes that there is an objective probability
measure $P$, often called ``real world'' or ``historical'' probability
measure. On the other hand, the absence of arbitrage implies the
existence of an equivalent martingale measure $P^*$, which should be
interpreted as a consistent price system that reflects the present ``market's belief''. From a mathematical point of view, the
coexistence of these measures and the explicit description of their
mutual densities is a rich source of technical exercises, and the
Girsanov transformation allows one to move freely back and forth
between $P$ and $P^*$. At a conceptual level, however, there is a
crucial difference between their roles.

The probability measure $P$ is usually seen as a probabilistic model
that tries to capture typical patterns observed in the past; under
implicit stationarity assumptions, it is then used as a forward-looking
prediction scheme. While it is often admitted that any specific choice
of $P$ involves a considerable amount of model risk, it is widely
believed that a true probability measure exists, and that probabilistic
models are getting better in approaching that reality.
Bruno de Finetti \cite{deFinetti1,deFinetti2}, however, would argue
that the problem is more fundamental than the issue of model risk. He
would put in doubt that it makes any sense to associate an objective
probability $P[A]$ to a financial event of the following type:
\[
A = \{\mbox{the sovereign bond with ISIN $x$ will not default} \}.
\]
On the other hand, a probability $P^*[A]$, or rather an expectation
$E^*[H]$ of the discounted future cash flow $H$ generated by the bond,
is assigned each day on the financial market, either directly through
the present market price of the bond or by the prices of instruments
such as credit default swaps (CDS) that provide insurance against a
default of the bond.
Thus the probability measure $P^*$ reflects the aggregate odds of a
large number of bets made on the market. This is in accordance with de
Finetti's claim that \textit{probability does not exist}, but that one
can of course take bets on a given event at certain odds. De Finetti
imposes consistency rules for the odds specified for different bets,
and he uses an emphatic ``you'' to stress the subjective nature of
the resulting probability measure~$P^*$. At the level of a single
agent, these
consistency rules may be viewed as an overly optimistic rationality
requirement. But if we replace de Finetti's ``you'' by ``the financial market'', this requirement becomes more compelling since
the market is more efficient in enforcing consistency via arbitrage
than any given individual. In fact, there is a close connection, both
at the conceptual and technical levels, between the fundamental theorem
of asset pricing and de Finetti's reconstruction of a probability
measure $P^*$ from a consistent system of bets; see, for example, \cite
{Mitter,Schervish}.

Apart from such foundational aspects, the attempts of predicting
financial developments in terms of an ``objective'' probability
measure $P$ can hardly been described as a success story, especially in
view of the recent financial crisis. On the other hand, a lot is known,
at any given time $t$, about the market's present predictions of
future developments in terms of a martingale measure $P^*_t$. More
precisely, the market's view at time $t$
is given by the conditional probability distribution
%
\begin{equation}
\label{P*tFt} P^*_t[ \cdot | \mathcal{F}_t] \quad \mbox{on}
\quad{\hat\mathcal{F}}_t
\end{equation}
where $\mathcal{F}_t$ is the $ \sigma$-field describing the available
information at time $t$, and ${\hat\mathcal{F}}_t$ is the
$ \sigma$-field generated by the pay-offs of traded contingent claims
with maturities $T > t$. Present prices of call or put options with
maturity $T$ provide information about the marginal distribution of
$P^*_t[  \cdot |  \mathcal{F}_t]$ at time $T>t$, and present prices
of more exotic options provide information about the multidimensional
marginals. This forward-looking ``lecture du march\'e'' is an
important part of current quantitative analysis.

At any given time $t$, the market's present view of the future as
expressed in the conditional pricing measure $P^*_t[  \cdot |
\mathcal{F}_t]$ is
consistent across different claims, and in particular it is
time-consistent across different maturities $T > t$.
But this consistent picture may change from day $t$ to day $t+1$, and
it may do so in a manner which is \textit{not} time-consistent.
Time-consistency across different dates $t$ may of course be desirable
from a normative point of view, and it is usually taken for granted in
the mathematical finance literature. In mathematical terms, it amounts
to the requirement that the conditional distributions in \eqref{P*tFt}
all belong to the same martingale measure $P^* \in\mathcal{P}^*$. In
the virtual world of a complete financial market model,
time-consistency would thus hold automatically, due to the fact that
the equivalent martingale measure is unique. In the larger world of
incomplete financial market models, and a fortiori in reality, one
should expect time-inconsistency. In our standard framework, this would
be described by a flow in the space $\mathcal{P}^*$ of
martingale measures. This flow could be continuous, but it also could
include jumps corresponding to abrupt regime changes.

Let us denote by $\mathcal{P}^*_{UI}$ the class of martingale measures
$P^*\in\mathcal{P}^*$ such that the price fluctuation $X$ is a
uniformly integrable martingale under $P^*$. Typically, both $\mathcal
{P}^*_{UI}$ and $\mathcal{P}^*_{NUI}:=\mathcal{P}^* \setminus\mathcal
{P}^*_{UI}$ are nonempty. The behavior of $X$ under a measure $P^*\in
\mathcal{P}^*_{NUI}$ is often interpreted as a bubble; cf.~\cite
{JarrowProtter1,JarrowProtter2}. A regime switch from an initial
martingale measure $P^*_0 \in\mathcal{P}^*_{UI}$, which does not
exhibit a bubble, to another martingale measure $P^*_1\in\mathcal
{P}^*_{NUI}$ would thus describe the sudden appearance of a bubble as
in \cite{JarrowProtter2}. But the flow in the space $\mathcal{P}^*$
could also move slowly from $P^*_0$ to $P^*_1$ as in \cite
{BiaginiFoellmer}, and this would induce the slow birth of a bubble as
a submartingale.

A deeper economic understanding of the dynamics of $P^*_t$ would
involve the microstructure of financial markets, i.e., the dynamic
behavior of agents with heterogeneous and interacting preferences and
expectations, with special emphasis on the ``herding'' effects
which are driving bubbles and crashes. So far, there are various toy
models, such as \cite{FoellmerKirman} and the references therein, which
try to capture some of these effects. But really compelling
microstructure models which offer serious possibilities for real-world
prediction are not yet in sight.

There is, however, an increasing need to complement the classical
microeconomic picture of noise traders and information traders
by taking into account a variety of trading algorithms which are
actually used on the financial market. In a way, this may make the
analysis of the resulting price dynamics more tractable, since the
structure of trading algorithms is more transparent and easier to model
than the behavioral characteristics of individual agents. While the
social utility of such algorithms may be debatable, it is important to
understand their effects as clearly as possible in mathematical terms.
In particular, such an understanding is crucial for any attempts to
design an intelligent regulatory framework that does not create new
arbitrage opportunities and thereby new sources of instability in the
financial system. In Sections~\ref{Price formation Section} and \ref
{Price impact Section} we are going to describe some of the simplest
mathematical issues which appear in connection with the interaction of
trading algorithms.

\section{Knightian uncertainty}

In recent years, there has been an increasing awareness, both among
practitioners and in academia, of the problems caused by an excessive
reliance on a specific probabilistic model and by the resulting ``control illusion''; see, for example, Section 4.9 in \cite{Hellwig}. As
a result, there is a renewed focus on the issue of \textit{model
uncertainty} or \textit{model ambiguity}, also called \textit{Knightian
uncertainty} in honor of Frank Knight \cite{Knight}, who introduced the
distinction between ``risk'' and ``uncertainty'' in the context
of economic decision theory. Here, ``risk'' refers to situations
where something is known about the probability measure $P$ (``known unknowns''), while ``uncertainty'' refers to situations where
this is not the case (``unknown unknowns''). In its analysis of the
recent subprime
crisis, the Turner Review \cite{Turner} distinguishes between ``mathematically modellable risk'' and Knightian uncertainty, and thus
seems to suggest that Knightian uncertainty is beyond the scope of
mathematical analysis. We do not share this conclusion. To the
contrary, we see Knightian uncertainty as a rich source of new
mathematical problems. This is illustrated by two recent developments,
where model uncertainty is taken into account explicitly. In Section
\ref{probability free Section}, we show how some key hedging arguments
in mathematical finance can be developed without even introducing any
probability measure. Another example is the specification of capital
requirements and of preferences in terms of convex risk measures,
described in Section~\ref{risk measure section}. Here the analysis is
not tied to the specific choice of a probability measure. Instead, one
considers a whole class of probabilistic models and takes a
conservative worst-case approach.

\subsection{Probability-free hedging}\label{probability free Section}

Consider a financial market with one risky and one riskless asset. In
mainstream finance, the price evolution of the risky asset is usually
modeled as a stochastic process defined on some probability space.
Here, however, we are going to work in a strictly pathwise setting. All
we assume is that the evolution of asset prices is given by one single
continuous non-negative trajectory $(X_t)_{0\le t\le T}$. As before, we will suppose
for simplicity that the prices of the riskless asset, or ``bond'',
are given by $B_t=1$ for all $t$.\vadjust{\goodbreak}

Now we discuss the possibility of dynamic trading in such a market.
To this end, consider a trading strategy $(\xi_t,\eta_t)_{0\le t\le
T}$, where $\xi_t$ describes the number of shares in the risky asset
and $\eta_t$ the number of shares in the bond held at time $t$. The
value of the portfolio $(\xi_t,\eta_t)$ is given by
%
\begin{equation}
\label{Vtdefinition} V_t=\xi_tX_t+
\eta_tB_t=\xi_tX_t+
\eta_t.
\end{equation}
To discuss investment or hedging strategies in this framework, it is
important to define self-financing trading strategies. Passing to the
continuous-time limit from a discrete-time framework suggests that the
strategy $(\xi_t,\eta_t)_{0\le t\le T}$ should be called self-financing
if the value process from~\eqref{Vtdefinition} satisfies the relation
%
\begin{equation}
\label{self-financing} V_t=V_0+\int_0^t
\xi_s \,dX_s,\qquad0\le t\le T,
\end{equation}
where the integral is the limit of nonanticipative Riemann sums:
%
\begin{equation}
\label{Ito integral} \int_0^t
\xi_s \,dX_s:=\lim_{n\uparrow\infty} \sum
_{t^n_i\le t}\xi_{t^n_{i-1}}(X_{t^n_i}-X_{t^n_{i-1}}).
\end{equation}
Here we can take for instance $t_i^n=i2^{-n}$. According to the results
in \cite{FoellmerIto}, this is possible when the trajectory $X$ admits
a continuous quadratic variation
\[
[X]_t=\lim_{n\uparrow\infty} \sum_{t^n_i\le
t}(X_{t^n_i}-X_{t^n_{i-1}})^2,
\qquad0\le t\le T,
\]
and if $\xi$ is of the form $\xi_t=g(X_t,A^1_t,\ldots, A^k_t)$ for a
continuous function $g$, which is differentiable in its first argument,
and for continuous trajectories\vspace*{1pt} $(A^i_t)_{0\le t\le T}$ of bounded variation.
In this case, it was shown in \cite{FoellmerIto} that It\^o's formula
holds for any $C^2$-function $f$ in the following strictly pathwise sense:
%
\begin{equation}
\label{Ito} f(X_t)=f(X_0)+\int_0^tf'(X_s)
\,dX_s+\frac12\int_0^tf''(X_s)
\,d[X]_s.
\end{equation}
Note that the second integral in \eqref{Ito} can be defined as a
classical Stieltjes integral, since $[X]_t$ is a nondecreasing function
of $t$.

As pointed out in \cite{FoellmerECM}, it follows immediately that a
non-constant trajectory $X$ must have nontrivial quadratic variation so
as to exclude arbitrage opportunities. Indeed, otherwise \eqref{Ito}
reduces to the standard fundamental theorem of calculus,
$f(X_t)=f(X_0)+\int_0^tf'(X_s) \,dX_s$, and by \eqref{self-financing}
the self-financing strategy $\xi_t=2(X_t-X_0)$ and $\eta_t=(X_t-X_0)^2-\xi_tX_t$
will generate the strictly positive wealth $V_t=(X_t-X_0)^2$ out of the
initial capital $V_0=0$.

The probability-free trading framework sketched above can for instance
be used to analyze the hedging error and the robustness of
model-specific hedging strategies such as in \cite{RobustBS} or \cite
{SchiedStadje}.\vadjust{\goodbreak} In some special cases, it is even possible to find
completely model-independent hedging strategies. We will illustrate
this now for the case of a \textit{variance swap} by transferring
arguments from \cite{Neuberger} and \cite{Dupire1993} to our
probability-free setting. A~{variance swap} is a path-dependent
financial derivative with payoff
\[
H=\sum_{i=1}^n(\log X_{t_{i+1}}-\log
X_{t_i})^2
\]
at time $T$, where $0<t_1<\cdots<t_n=T$ are fixed time points. These
time points are often chosen so that $X_{t_i}$ is the {closing price}
of the risky asset at the end of the $i$th trading day; see,
e.g., \cite{BuehlerThesis,Buehler,Gatheralbook} for background on
variance swaps.
When $n$ is large enough, the payoff of the variance swap can thus be
approximated by the quadratic variation of $\log X$, i.e.,
%
\begin{equation}
\label{swap} H\approx[\log X]_T=\int_0^T
\frac1{X^2_t} \,d[X]_t.
\end{equation}
Here, the second identity follows, e.g., from Proposition 2.2.10 in
\cite{Sondermann}. On the other hand, applying It\^o's formula \eqref
{Ito} to the function $f(x)=\log x$ yields
%
\begin{equation}
\label{logItoEq} \log X_T-\log X_0=\int
_0^T\frac1{X_t} \,dX_t-
\frac12\int_0^T\frac1{X_t^2}
\,d[X]_t.
\end{equation}
Putting \eqref{swap} and \eqref{logItoEq} together implies that
%
\begin{eqnarray}
\label{h} H\approx\int_0^T
\frac1{X_t^2} \,d[X]_t=2\log
X_0-2\log X_T+2\int_0^T
\frac1{X_t} \,dX_t.
\end{eqnarray}
The It\^o integral on the right-hand side of \eqref{h} can be regarded
as the terminal value of the self-financing trading strategy that has
zero initial investment and otherwise consists in holding $\xi_t=2/X_t$
shares of the risky asset at each time $t$. To interpret the two
logarithmic terms in \eqref{h}, we apply the Breeden--Litzenberger formula,
%
\begin{eqnarray}
\label{BreedenLitzenberger} h(X_T)&=&h(X_0)+h^\prime(X_0)
(X_T-X_0)+\int_{0}^{X_0}(K-X_T)^+
h''(K) \,dK
\nonumber
\\[-8pt]
\\[-8pt]
\nonumber
&&{}+ \int_{X_0}^{\infty}(X_T-K)^+
h''(K) \,dK
\end{eqnarray}
(e.g., \cite{FoellmerSchied}, Exercise 1.3.3) to the function
$h(x)=\log x$ and obtain
%
\begin{eqnarray}
\label{ModelindependentVarSwapHedgeEq} H&\approx&-\frac2{X_0}(X_T-X_0)+
\int_0^{X_0}(K-X_T)^+
\frac2{K^2} \,dK+ \int_{X_0}^\infty(X_T-K)^+
\frac2{K^2} \,dK
\nonumber
\\[-8pt]
\\[-8pt]
\nonumber
&&{}+2\int_0^T
\frac1{X_t} \,dX_t.
\end{eqnarray}
That is, $H$ can be hedged by selling $2/X_0$ zero-price forward
contracts, holding portfolios consisting of $2/K^2 \,dK$ ``out-of-the-money'' put\vadjust{\goodbreak}
and call options with maturity $T$ for each strike $K$, and using the
self-financing trading strategy with
$\xi_t=2/X_t$. The most remarkable feature of this hedging strategy is
that it is {model-independent}. That is, \eqref
{ModelindependentVarSwapHedgeEq} is valid independently of possible
probabilistic dynamics of the price process $X$. The hedging strategy
is therefore not subject to {model risk} that might result from a
misspecification of such probabilistic dynamics.

Similar results as obtained for variance swaps are valid for so-called
\textit{Gamma} or \textit{entropy swaps} with payoff
\[
\sum_{i=1}^nX_{t_i}(\log
X_{t_{i+1}}-\log X_{t_i})^2
\]
and also for \textit{corridor variance swaps} with payoff
\[
\sum_{i=1}^n{\bm1}_{\{A\le X_{t_i}\le B\}}(\log
X_{t_{i+1}}-\log X_{t_i})^2,
\]
for some real numbers $A, B$, with $A<B$. See also \cite{DavisObloj}
for further extensions.

Note that the Breeden--Litzenberger formula \eqref{BreedenLitzenberger}
can be regarded as a simple static, and hence model-free, hedge for the
option $h(X_T)$ in terms of standard ``plain vanilla'' put and
call options. In some cases, static hedges (or superhedges) can also be constructed for
path-dependent derivatives such as barrier or lookback options; see,
e.g., \cite{BrownHobsonRogers,CoxObloj,Hobson}.

If uncertainty is restricted to a suitable class of scenarios, the
strictly pathwise approach can also be used to formulate the crucial
hedging argument of Section~\ref{Perfect hedge Section} in a
probability-free manner. To this end, we fix a continuous volatility
function $\sigma(x,t)>0$ on $[0,\infty)\times[0,T]$ and restrict the
possible scenarios to the set $\Omega_\sigma$ of all nonnegative
continuous functions $\omega$ on $[0,T]$ such that the coordinate
process $X_t(\omega)=\omega(t)$ admits an absolutely continuous
quadratic variation $d[X(\omega)]_t=\sigma^2(X_t(\omega),t)X_t^2(\omega
) \,dt$. Consider a derivative of the form $H=h(X_T)$. As explained in
\cite{BickWillinger} or \cite{FoellmerECM}, we can now use the
time-dependent extension of the pathwise It\^o formula \eqref{Ito} to
construct a perfect hedge of the form $\xi_t(\omega)=F_x(X_t(\omega
),t)$, where $F$ solves an appropriate parabolic equation with boundary
condition $F(x,T)=h(x)$.
Moreover, a theorem of Paul L\'evy implies that there is exactly one
probability measure $P^*$ on the space $\Omega_\sigma$ such that the
coordinate process $X$ becomes a martingale under $P^*$. The price of
the derivative $H$, defined as the initial cost of the perfect hedge,
can then be computed as in \eqref{unique price} as the expected value
$E^*[H]$ of $H$ under the measure $P^*$.

In order to extend the preceding construction to more exotic options,
one can use a strictly pathwise version of Malliavin calculus as
recently developed in \cite{Dupire} and \cite{ContFournier}. For an
alternative pathwise approach in terms of \textit{rough paths,} see \cite
{LyonsQian,FrizVictoir}.

\subsection{Monetary risk measures}\label{risk measure section}

The \textit{capital requirement} associated with the profits and losses,
or P\&L, of a given financial position is specified as the minimal
capital that should be added to the position in order to make that
position acceptable from the point of view of a supervising agency.
This idea can be formalized as follows by the notion of a \emph
{monetary measure of risk}.\vadjust{\goodbreak}

The P\&L describes the uncertain net monetary outcome at the end of a
given trading period, and so it will be modeled as a real-valued
measurable function $X$ on a measurable space $(\Omega,\mathcal{F})$ of
possible scenarios. We fix a linear space $\mathcal{X}$ of such P\&Ls
and a nonempty subset $\mathcal{A}\subset\mathcal{X}$ associated with
those positions that are deemed acceptable. We require that $\mathcal{X}$ contains all constants and that $Y\in
\mathcal{A}$ whenever $Y\geq X$ for some $X\in\mathcal{A}$. The
functional $\rho$ on $\mathcal{X}$ defined by
%
\begin{equation}
\label{monetary rm} \rho(X):=\inf\{m\in\mathbb{R} | X+m\in\mathcal{A}\}
\end{equation}
is then called a \textit{monetary risk measure}, and the value $\rho(X)$
is interpreted as the capital requirement for the financial position
with P\&L $X$.

The standard example of a monetary risk measure is \textit{Value at Risk}
at some level $\lambda\in(0,1)$. For a given probabilistic model
described by a probability measure $P$ on $(\Omega,\mathcal{F})$, $X$
is deemed acceptable for Value at Risk if the probability $P[X<0]$ of a
shortfall does not exceed the level~$\lambda$. The resulting monetary
risk measure \eqref{monetary rm} is then given, up to a minus sign, by
a $\lambda$-quantile of the distribution of $X$ under $P$. Value at
Risk is widely used in practice. But it also has a number of
deficiencies. In particular, it does not account for the size of a
possible shortfall and therefore penalizes diversification while
encouraging the concentration of risk. The recognition of these
deficiencies motivated the axiomatic approach to a general theory of
monetary risk measures as initiated by Artzner, Delbaen, Eber, and
Heath \cite{ADEH} in the late nineties. But there are also other
drawbacks. For instance, in reaction to the recent financial crisis,
\textit{The Turner Review -- A~regulatory response to the global banking
crisis} \cite{Turner} emphasizes an excessive reliance on a single
probabilistic model $P$ and thus raises the issue of Knightian uncertainty.

We are now going to sketch some of the key ingredients in the theory of
convex risk measures. As we will see, this theory does not only address
the issue that diversification should not be penalized by the capital
requirement. It also provides a case study on how to deal with
Knightian uncertainty in a mathematical framework.

To capture the idea that diversification should be encouraged rather
than penalized by a monetary risk measure, we require that the
acceptance set $\mathcal{A}$ be convex. In this case, the monetary risk
measure $\rho$ defined via \eqref{monetary rm} is called a \textit{convex
risk measure}, because convexity of $\mathcal{A}$ is equivalent to
convexity of $\rho$. When $\mathcal{A}$ is even a convex cone, $\rho$
is called a \textit{coherent} risk measure. The notion of a coherent risk
measures was introduced in the seminal paper \cite{ADEH}; the
subsequent extension from coherent to convex risk measures was
introduced independently in \cite{Heath}, \cite
{FrittelliRosazzaGianin}, and \cite{FoellmerSchiedConv}.
Convex duality implies that a convex risk measure typically takes the form
%
\begin{equation}
\label{risk measure rep} \rho(X)=\sup_{Q\in\mathcal{Q}_\rho}\bigl\{E_Q[-X]-
\alpha(Q)\bigr\},
\end{equation}
where $\mathcal{Q}_\rho$ is some class of probability measures and
$\alpha:\mathcal{Q}_\rho\to{\mathbb R}\cup\{+\infty\}$ is a penalty
function. The capital requirement is thus determined as follows: The
expected loss of a position is calculated for each probability measure
$Q\in\mathcal{Q}_\rho$ and penalized by the penalty $\alpha(Q)$; then
one takes the worst penalized expected loss over the class $\mathcal
{Q}_\rho$.
This procedure can be interpreted as follows in the light of model uncertainty.
No probability measure is fixed in advance, but probability measures
$Q\in\mathcal{Q}_\rho$ do come in via convex duality and take the role
of stress tests.
The set $\mathcal{Q}_\rho$ can be regarded as a class of plausible
probabilistic models, in which each model $Q\in\mathcal{Q}_\rho$ is
taken more or less seriously, according to the size of the penalty
$\alpha(Q)$. In this way, model uncertainty is taken into account
explicitly. In the special coherent case the penalty function will
vanish on~$\mathcal{Q}_\rho$, and so the representation \eqref{risk
measure rep} reduces to
%
\begin{equation}
\label{coh representation} \rho(X)=\sup_{Q\in\mathcal{Q}_\rho}E_Q[-X],
\end{equation}
that is, to the worst case expected loss over the class $\mathcal
{Q}_\rho$.

In the context of an arbitrage-free but possibly incomplete financial
market model, the superhedging risk measure,
\[
\rho(X) = \sup_{P^*\in\mathcal{P}^*} E^* [-X] ,
\]
is clearly a coherent risk measure. The corresponding acceptance set
$\mathcal A$ consists of all $X$ for which one can find a dynamic
trading strategy with initial capital $V_0=0$ and final outcome $V_T$
such that the pay-off of the combined position
$X+ V_T$ is nonnegative with probability one.

In the setting of mathematical finance, the history of coherent and
convex risk measures begins with the seminal paper \cite{ADEH}, as
mentioned above. In a wider mathematical context, however,
there is a considerable pre-history in areas such as in game theory and
Choquet integration \cite{Dellacherie,Schmeidler}, robust statistics
\cite{Huber,HuberStrassen}, and actuarial premium principles \cite
{DeprezGerber,Goovaerts}.

Risk measures have also appeared implicitly in the microeconomic theory
of preferences. Preferences on the space $\mathcal{X}$ are usually
represented by some utility functional $U$ on $\mathcal{X}$. Under the
axioms of rationality as formulated by von Neumann and Morgenstern \cite
{NeumannMorgenstern} and Savage \cite{Savage}, $U$~takes the form of an
\textit{expected utility}, i.e.,
%
\begin{equation}
\label{expected utility} U(X)=E_P\bigl[u(X)\bigr]
\end{equation}
for some increasing continuous function $u$ and some probability
measure $P$ on $(\Omega,\mathcal{F})$. As shown by Gilboa and
Schmeidler \cite{GilboaSchmeidler} in the late eighties, a natural
relaxation of the axioms of rationality implies that the linear risk
measure $-E_P[ \cdot ]$ in \eqref{expected utility} should be
replaced by a general coherent risk measure $\rho$:
\[
U(X)=-\rho\bigl(u(X)\bigr)=\inf_{Q\in\mathcal{Q}_\rho}E_Q\bigl[u(X)\bigr].
\]
More recently, Maccheroni, Marinacci, and Rustichini \cite
{Maccheronietal} have relaxed the rationality axioms even further. In
their axiomatic setting, $\rho$ is now a convex risk measure, and so
the numerical representation of preferences takes the form
\[
U(X)=-\rho\bigl(u(X)\bigr)=\inf_{Q\in\mathcal{Q}_\rho}\bigl\{E_Q\bigl[u(X)
\bigr]+\alpha(Q)\bigr\}.
\]
While classical risk aversion is captured by concavity of the utility
function $u$, the concavity of $-\rho$
corresponds to a behavioral assumption of model uncertainty aversion;
see \cite{GilboaSchmeidler}, \cite{Maccheronietal}, and also \cite
{FoellmerSchied}.

\section{Price formation, market microstructure, and the emergence of
algorithmic trading}\label{Price formation Section}

When L. Bachelier and P.A. Samuelson formulated their models of asset
price processes, orders were usually executed by broker signals in
trading pits. But in recent years, the way in which financial markets
operate has changed dramatically. We are now going to discuss some of
the new challenges for mathematical finance that are resulting from
this change.

In 1971, the world's first electronic stock exchange, NASDAQ, was
opened. In the subsequent decades, fostered by measures of market
deregulation and technological improvements, more and more trading pits
were abandoned and replaced by fully electronic exchanges. Such an
electronic exchange basically operates with two different kinds of
orders, \textit{limit orders} and \textit{market orders}.
A~limit order is an order to buy or sell a certain amount of shares at
a specific price. It is collected in an electronic \textit{limit order book}
until there is a matching sell or buy order. A market order is an order
to buy or sell a certain amount of shares at the best currently
available price. It thus consumes limit orders according to price
priority. When the total size of all limit orders at the best price is
larger than the size of the incoming matching order, limit orders are
usually executed according to a first-in first-out rule. On this
microscopic level, asset price dynamics are thus represented not by a
one-dimensional diffusion process but by the evolution of the entire
limit order book, which, from a mathematical point of view, can be
regarded as a complex queuing system. As such, it can at least in
principle be modeled mathematically.
With a suitable model at hand, one can try to ``zoom out'' of
the microscopic picture and characterize the limiting dynamics of the
mid price (i.e., the average between the best buy and sell limit
orders) on a mesoscopic diffusion scale. This can either lead to a
confirmation of the standard modeling paradigms of mathematical finance
or to the discovery of new types of asset price dynamics. Initial
studies concerned with such questions were conducted in \cite
{AvellanedaStoikov,Bovier,ContDeLarrard1,ContDeLarrard2,ContKukanovStoikov}
with, e.g., \cite{ContDeLarrard2} finding a Bachelier-type model in the
diffusion limit.

The emergence of electronic trading venues facilitated the use of
computers for order placement, and soon the new phenomena of \emph
{algorithmic} and \textit{high-frequency trading} came into existence.
Today, limit order books are updated in time intervals measured in
milliseconds so that no human being can possibly follow
the price evolution of sufficiently liquid assets. The use of computers
is hence mandatory for market makers and many other traders. As a
consequence, the
vast majority of orders in equity markets is now placed by computer
algorithms. A~good description of the current state of electronic
markets is given in \cite{Lehalle}.

The computerization of financial markets led to some effects that can
be regarded as potentially beneficial. For instance, the liquidity
provided by high-frequency market makers and the competition between
the growing number of electronic trading venues contributed to a
significant decline of bid-ask spreads, thus reducing transaction costs
for ordinary investors.
There was also some hope that computer programs would act more
rationally than human investors, in particular in critical situations,
and thus avoid panic and herding behavior.
These hopes, however, were seriously challenged by the \textit{Flash
Crash} of May 6, 2010. On that day, a sell order placed in a nervous
market triggered a ``hot-potato
game'' among the trading algorithms of high-frequency traders (HFTs), which resulted in the steepest drop of asset prices ever,
followed by a sharp recovery within 20 minutes.
The following quote from \cite[page 3]{SEC} gives some indication
that the Flash Crash was indeed generated by a feedback overflow
between several trading algorithms:

\begin{quote}
\ldots  HFTs began to quickly buy and then
resell contracts to each other -- generating a ``hot-potato'' volume
effect as the same positions were rapidly passed back and forth.
Between 2:45:13 and 2:45:27, HFTs traded over 27,000 contracts, which
accounted for about 49 percent of the total trading volume, while
buying only about 200 additional contracts net.
\end{quote}

\noindent It is an interesting challenge to understand the reasons why
interacting trading algorithms can end up in such a ``hot-potato
game'' and to reproduce this phenomenon in a mathematical model.
As we will see in the next section, there are already some preliminary
results that may be connected to this phenomenon.

Besides the possible creation of crash scenarios, there are also other
aspects of electronic trading that are potentially problematic. For
instance, certain predatory trading algorithms scan order signals for
patterns resulting from the execution of large trades. Once such a
large trade is detected, the predatory trading algorithm tries to make
a profit by building up a position whose value will be increased by the
price impact generated by the large trade; see \cite
{BrunnermeierPedersen,Carlinetal,SchoenebornSchied}.
To escape the adverse effects of price impact and predatory trading,
many investors resort to so-called dark pools, in which orders are
invisible to other market participants. But the fact that many dark
pools derive the execution price of orders from the `lit' market
facilitates predatory trading techniques such as `fishing', which
are based on manipulating the price in the lit market; see \cite
{KSS,KratzSchoeneborn,Mittal}.

\section{Price impact and order execution}\label{Price impact Section}

The key to understanding algorithmic trading and its potential benefits
and risks is the phenomenon of \textit{price impact}, i.e., the fact that
the execution of a large order influences the price of the underlying
asset. It is one of the basic mechanisms by which economic agents interact with
the market and, thus, with each other. Spectacular cases in which price
impact played an important role were the debacle of Metallgesellschaft
in 1993, the LTCM crisis in 1998, or the unwinding of J\'er\^ome
Kerviel's portfolio by Societ\'e G\'en\'erale in 2008. But price
impact can also be significant in much smaller trades, and it belongs
to the daily business of many financial institutions.

The first step in understanding price impact is the execution of a
single trade, a problem at which one can look on several scales. On a
microscopic scale, one considers a trade that is small enough to be
executed by placing a single order in a limit order book. When this
order is placed as a market order, it will impact the limit order book
by consuming limit orders and, if it is large enough, shift the
corresponding best price and widen the bid-ask spread; see \cite
{AFS2,ow,WeberRosenow}. When it consists in placing or cancelling a
limit order, its quantitative impact is not as easy to describe but it
is nevertheless existing. In either case, the impact of a trade is
transient and will eventually diminish, a fact that becomes important
on the next, mesoscopic level.

Many trades are too big to be executed in one single order and
therefore need to be split in a series of smaller orders, sometimes
called `child orders', which are then spread out over a certain
time interval. On this mesoscopic scale, trading algorithms are used to
determine sizes and timing of each child order. These algorithms are
typically based on a market impact model, i.e., a stochastic model for
asset prices that takes into account the feedback effects of trading strategies.
We refer to \cite{GatheralSchiedSurvey} for a survey on some
models that are currently available. The problem of determining \emph
{optimal} trade execution strategies for a given cost criterion in a
specific model has a rich structure and often leads to questions that
are of intrinsic mathematical interest. It is, for instance, connected
to the mathematical topics of finite-fuel control, Choquet capacity
theory, and Dawson--Watanabe superprocesses. Let us briefly sketch the
latter connection as established in \cite{SchiedFuel}. When formulating
the optimal trade execution problem as a stochastic control problem,
the liquidation constraint translates into a singular terminal
condition for the corresponding Hamilton--Jacobi--Bellman equation.
This equation can be further reduced to a quasilinear parabolic partial
differential equation with infinite terminal condition. But, according
to \cite{DynkinParabolic}, such equations are related to the Laplace
functionals of Dawson--Watanabe superprocesses.

The existence or nonexistence and the
structure of optimal trade execution strategies can also yield
information on the viability of the underlying market impact model and
perhaps even on the nature of price impact itself; see, e.g., \cite
{ASS,Gatheral,HubermanStanzl}. For instance, it was shown in \cite
{ASS} that the price impact of single orders must decay as a convex
function of time to exclude oscillatory trade execution strategies that
are to some extend reminiscent of the ``hot-potato
game'' mentioned earlier.

It should be pointed out that the market impact models currently
available in the literature are all relatively simple. In particular,
there is yet no
model that combines both transience and nonlinearity of price impact in
a truly convincing way.

On a macroscopic scale, the execution of the trade is seen in relation
to the behavior of other agents---or algorithms---in the market. As
mentioned above, the fact that an agent is executing a large trade can
be betrayed to competitors for instance via the order signals created
by the execution algorithm. When a competitor detects the execution of
a large trade, it is generally believed that predatory trading, as
described above, is the corresponding profit-maximizing strategy. This
was also obtained as a mathematical result in \cite{Carlinetal} by
analyzing a game-theoretic setting. By slightly extending this setting,
however, it was found in \cite{SchoenebornSchied} that predatory
trading may become suboptimal in markets that are sufficiently `elastic'
in the sense that the price impact of orders decays very
rapidly. In such markets it is instead beneficial for the competitor to
cooperate with the large trader and to provide liquidity. A completely
different pattern occurs, however, when price impact is transient. Sch\"
oneborn \cite{Schoeneborn} showed that in a discrete-time model with
linear, exponentially decaying price impact the large trader and the
competitor start a ``hot-potato
game'' very similar to the one observed in the Flash Crash.
%


\printhistory


\begin{thebibliography}{108}

\bibitem{AFS2}
%
\begin{barticle}[mr]
\bauthor{\bsnm{Alfonsi},~\bfnm{Aur{\'e}lien}\binits{A.}},
\bauthor{\bsnm{Fruth},~\bfnm{Antje}\binits{A.}} \AND
\bauthor{\bsnm{Schied},~\bfnm{Alexander}\binits{A.}}
(\byear{2010}).
\btitle{Optimal execution strategies in limit order books with general shape
functions}.
\bjournal{Quant. Finance}
\bvolume{10}
\bpages{143--157}.
\bid{doi={10.1080/14697680802595700}, issn={1469-7688}, mr={2642960}}
\bptok{imsref}%
\end{barticle}
%
\endbibitem

\bibitem{ASS}
%
\begin{barticle}[author]
\bauthor{\bsnm{Alfonsi},~\bfnm{Aur{\'e}lien}\binits{A.}},
\bauthor{\bsnm{Schied},~\bfnm{Alexander}\binits{A.}} \AND
\bauthor{\bsnm{Slynko},~\bfnm{Alla}\binits{A.}}
(\byear{2012}).
\btitle{Order book resilience, price manipulation, and the positive portfolio
problem}.
\bjournal{SIAM J. Financial Math.}
\bvolume{3}
\bpages{511--533}.
\bid{mr={2968044}}
\bptok{imsref}%
\end{barticle}
%
\endbibitem

\bibitem{ADEH}
%
\begin{barticle}[mr]
\bauthor{\bsnm{Artzner},~\bfnm{Philippe}\binits{P.}},
\bauthor{\bsnm{Delbaen},~\bfnm{Freddy}\binits{F.}},
\bauthor{\bsnm{Eber},~\bfnm{Jean-Marc}\binits{J.M.}} \AND
\bauthor{\bsnm{Heath},~\bfnm{David}\binits{D.}}
(\byear{1999}).
\btitle{Coherent measures of risk}.
\bjournal{Math. Finance}
\bvolume{9}
\bpages{203--228}.
\bid{doi={10.1111/1467-9965.00068}, issn={0960-1627}, mr={1850791}}
\bptok{imsref}%
\end{barticle}
%
\endbibitem

\bibitem{AvellanedaStoikov}
%
\begin{barticle}[mr]
\bauthor{\bsnm{Avellaneda},~\bfnm{Marco}\binits{M.}} \AND
\bauthor{\bsnm{Stoikov},~\bfnm{Sasha}\binits{S.}}
(\byear{2008}).
\btitle{High-frequency trading in a limit order book}.
\bjournal{Quant. Finance}
\bvolume{8}
\bpages{217--224}.
\bid{doi={10.1080/14697680701381228}, issn={1469-7688}, mr={2408299}}
\bptok{imsref}%
\end{barticle}
%
\endbibitem

\bibitem{Bachelier}
%
\begin{bbook}[mr]
\bauthor{\bsnm{Bachelier},~\bfnm{Louis}\binits{L.}}
(\byear{1995}).
\btitle{Th\'eorie de la Sp\'eculation: Th{\'e}orie Math{\'e}matique du Jeu}.
\bseries{Les Grands Classiques Gauthier-Villars. [Gauthier-Villars Great
Classics]}.
\blocation{Sceaux}: \bpublisher{\'Editions Jacques Gabay}.
\bnote{Reprint of the 1900 original}.
\bid{mr={1397712}}
\bptok{imsref}%
\end{bbook}
%
\endbibitem

\bibitem{Bernoulli}
%
\begin{barticle}[author]
\bauthor{\bsnm{Bernoulli},~\bfnm{Daniel}\binits{D.}}
(\byear{1738}).
\btitle{Specimen theoriae novae de mensura sortis}.
\bjournal{Commentarii Academiae Scientiarum Imperialis Petropolitanae}
\bvolume{5}
\bpages{175--1926}.
\bnote{Translated by L. Sommer: \textit{Econometrica} \textbf{22}
(1954) 23--36}.
\bptok{imsref}%
\end{barticle}
%
\endbibitem

\bibitem{BiaginiFoellmer}
%
\begin{bmisc}[author]
\bauthor{\bsnm{Biagini},~\bfnm{Francesca}\binits{F.}},
\bauthor{\bsnm{F{\"o}llmer},~\bfnm{Hans}\binits{H.}} \AND
\bauthor{\bsnm{Nedelcu},~\bfnm{S.}\binits{S.}}
(\byear{2011}).
\bhowpublished{Shifting martingale measures and the birth of a bubble as a submartingale. Unpublished manuscript.}
\bptok{imsref}%
\end{bmisc}
%
\endbibitem

\bibitem{BickWillinger}
%
\begin{barticle}[mr]
\bauthor{\bsnm{Bick},~\bfnm{Avi}\binits{A.}} \AND
\bauthor{\bsnm{Willinger},~\bfnm{Walter}\binits{W.}}
(\byear{1994}).
\btitle{Dynamic spanning without probabilities}.
\bjournal{Stochastic Process. Appl.}
\bvolume{50}
\bpages{349--374}.
\bid{doi={10.1016/0304-4149(94)90128-7}, issn={0304-4149}, mr={1273780}}
\bptok{imsref}%
\end{barticle}
%
\endbibitem

\bibitem{BlackScholes}
%
\begin{barticle}[author]
\bauthor{\bsnm{Black},~\bfnm{Fischer}\binits{F.}} \AND
\bauthor{\bsnm{Scholes},~\bfnm{Myron}\binits{M.}}
(\byear{1973}).
\btitle{The pricing of options and corporate liabilities}.
\bjournal{The Journal of Political Economy}
\bpages{637--654}.
\bptok{imsref}%
\end{barticle}
%
\endbibitem

\bibitem{Mitter}
%
\begin{barticle}[mr]
\bauthor{\bsnm{Borkar},~\bfnm{V.~S.}\binits{V.S.}},
\bauthor{\bsnm{Konda},~\bfnm{V.~R.}\binits{V.R.}} \AND
\bauthor{\bsnm{Mitter},~\bfnm{S.~K.}\binits{S.K.}}
(\byear{2004}).
\btitle{On {D}e {F}inetti coherence and {K}olmogorov probability}.
\bjournal{Statist. Probab. Lett.}
\bvolume{66}
\bpages{417--421}.
\bid{doi={10.1016/j.spl.2003.11.011}, issn={0167-7152}, mr={2045135}}
\bptok{imsref}%
\end{barticle}
%
\endbibitem

\bibitem{Bovier}
%
\begin{barticle}[mr]
\bauthor{\bsnm{Bovier},~\bfnm{Anton}\binits{A.}},
\bauthor{\bsnm{{\v{C}}ern{\'y}},~\bfnm{Ji{\v{r}}{\'{\i}}}\binits{J.}}
\AND
\bauthor{\bsnm{Hryniv},~\bfnm{Ostap}\binits{O.}}
(\byear{2006}).
\btitle{The opinion game: Stock price evolution from microscopic market
modeling}.
\bjournal{Int. J. Theor. Appl. Finance}
\bvolume{9}
\bpages{91--111}.
\bid{doi={10.1142/S0219024906003421}, issn={0219-0249}, mr={2205716}}
\bptok{imsref}%
\end{barticle}
%
\endbibitem

\bibitem{BrownHobsonRogers}
%
\begin{barticle}[mr]
\bauthor{\bsnm{Brown},~\bfnm{Haydyn}\binits{H.}},
\bauthor{\bsnm{Hobson},~\bfnm{David}\binits{D.}} \AND
\bauthor{\bsnm{Rogers},~\bfnm{L.~C.~G.}\binits{L.C.G.}}
(\byear{2001}).
\btitle{Robust hedging of barrier options}.
\bjournal{Math. Finance}
\bvolume{11}
\bpages{285--314}.
\bid{doi={10.1111/1467-9965.00116}, issn={0960-1627}, mr={1839367}}
\bptok{imsref}%
\end{barticle}
%
\endbibitem

\bibitem{BrunnermeierPedersen}
%
\begin{barticle}[author]
\bauthor{\bsnm{Brunnermeier},~\bfnm{Markus~K.}\binits{M.K.}} \AND
\bauthor{\bsnm{Pedersen},~\bfnm{Lasse~Heje}\binits{L.H.}}
(\byear{2005}).
\btitle{Predatory trading}.
\bjournal{J. Finance}
\bvolume{60}
\bpages{1825--1863}.
\bptok{imsref}%
\end{barticle}
%
\endbibitem

\bibitem{Buehler}
%
\begin{barticle}[mr]
\bauthor{\bsnm{B{\"u}hler},~\bfnm{Hans}\binits{H.}}
(\byear{2006}).
\btitle{Consistent variance curve models}.
\bjournal{Finance Stoch.}
\bvolume{10}
\bpages{178--203}.
\bid{doi={10.1007/s00780-006-0008-2}, issn={0949-2984}, mr={2223095}}
\bptok{imsref}%
\end{barticle}
%
\endbibitem

\bibitem{BuehlerThesis}
%
\begin{bmisc}[author]
\bauthor{\bsnm{B{\"u}hler},~\bfnm{Hans}\binits{H.}}
(\byear{2006}).
\bhowpublished{Volatility markets: Consistent modeling, hedging and practical
implementation. Ph.D. thesis, TU Berlin.}
\bptok{imsref}%
\end{bmisc}
%
\endbibitem

\bibitem{Carlinetal}
%
\begin{barticle}[author]
\bauthor{\bsnm{Carlin},~\bfnm{Bruce~Ian}\binits{B.I.}},
\bauthor{\bsnm{Lobo},~\bfnm{Miguel~Sousa}\binits{M.S.}} \AND
\bauthor{\bsnm{Viswanathan},~\bfnm{S.}\binits{S.}}
(\byear{2007}).
\btitle{Episodic liquidity crises: Cooperative and predatory trading}.
\bjournal{J. Finance}
\bvolume{65}
\bpages{2235--2274}.
\bptok{imsref}%
\end{barticle}
%
\endbibitem

\bibitem{Cassidy}
%
\begin{bbook}[author]
\bauthor{\bsnm{Cassidy},~\bfnm{J.}\binits{J.}}
(\byear{2009}).
\btitle{How Markets Fail: The Logic of Economic Calamities}.
\blocation{New York}: \bpublisher{Farrar, Straus~\& Giroux}.
\bptok{imsref}%
\end{bbook}
%
\endbibitem

\bibitem{SEC}
%
\begin{bmisc}[author]
\borganization{CFTC-SEC}.
(\byear{2010}).
\bhowpublished{Findings regarding the market events of {M}ay 6, 2010. Technical
report.}
\bptok{imsref}%
\end{bmisc}
%
\endbibitem

\bibitem{ContDeLarrard2}
%
\begin{bmisc}[author]
\bauthor{\bsnm{Cont},~\bfnm{Rama}\binits{R.}} \AND\bauthor{\bparticle{de}
\bsnm{Larrard},~\bfnm{Adrien}\binits{A.}}
(\byear{2010}).
\bhowpublished{Linking volatility with order flow: Heavy traffic
approximations and
diffusion limits of order book dynamics. Unpublished manuscript.}
\bptok{imsref}%
\end{bmisc}
%
\endbibitem

\bibitem{ContDeLarrard1}
%
\begin{bmisc}[author]
\bauthor{\bsnm{Cont},~\bfnm{Rama}\binits{R.}} \AND\bauthor{\bparticle{de}
\bsnm{Larrard},~\bfnm{Adrien}\binits{A.}}
(\byear{2013}).
\bhowpublished{Price dynamics in a {M}arkovian limit order market. \textit{SIAM J. Financial Math.} To appear.}
\bptok{imsref}%
\end{bmisc}
%
\endbibitem

\bibitem{ContFournier}
%
\begin{bmisc}[author]
\bauthor{\bsnm{Cont},~\bfnm{R.}\binits{R.}} \AND
\bauthor{\bsnm{Fournie},~\bfnm{D.~A.}\binits{D.A.}}
(\byear{2013}).
\bhowpublished{Functional Ito calculus and stochastic integral representation
of martingales. \textit{Ann. Probab.} \textbf{41} 109--133.}
\bptok{imsref}%
\end{bmisc}
%
\endbibitem

\bibitem{ContKukanovStoikov}
%
\begin{bmisc}[author]
\bauthor{\bsnm{Cont},~\bfnm{Rama}\binits{R.}},
\bauthor{\bsnm{Kukanov},~\bfnm{Arseniy}\binits{A.}} \AND
\bauthor{\bsnm{Stoikov},~\bfnm{Sasha}\binits{S.}}
(\byear{2010}).
\bhowpublished{The price impact of order book events. Unpublished manuscript. Available at \arxivurl{arXiv:1011.6402}.}
\bptok{imsref}%
\end{bmisc}
%
\endbibitem


\bibitem{CoxObloj}
%
\begin{barticle}[mr]
\bauthor{\bsnm{Cox},~\bfnm{A.~M.~G.}\binits{A.M.G.}} \AND
\bauthor{\bsnm{Ob{\l}{\'o}j},~\bfnm{Jan}\binits{J.}}
(\byear{2011}).
\btitle{Robust hedging of double touch barrier options}.
\bjournal{SIAM J. Financial Math.}
\bvolume{2}
\bpages{141--182}.
\bid{doi={10.1137/090777487}, issn={1945-497X}, mr={2772387}}
\bptok{imsref}%
\end{barticle}
%
\endbibitem

\bibitem{DavisObloj}
%
\begin{bmisc}[author]
\bauthor{\bsnm{Davis},~\bfnm{Mark}\binits{M.}},
\bauthor{\bsnm{Ob{\l}{\'o}},~\bfnm{Jan}\binits{J.}} \AND
\bauthor{\bsnm{Raval},~\bfnm{Vimal}\binits{V.}}
(\byear{2013}).
\bhowpublished{Arbitrage bounds for weighted variance swap prices.
\textit{Math. Finance}. To appear.}
\bptok{imsref}%
\end{bmisc}
%
\endbibitem

\bibitem{deFinetti1}
%
\begin{bbook}[mr]
\bauthor{\bparticle{de} \bsnm{Finetti},~\bfnm{Bruno}\binits{B.}}
(\byear{1990}).
\btitle{Theory of Probability: A Critical Introductory Treatment.
{V}ol. 1}.
\bseries{Wiley Classics Library}.
\blocation{Chichester}: \bpublisher{Wiley}.
\bid{mr={1093666}}
\bptok{imsref}%
\end{bbook}
%
\endbibitem

\bibitem{deFinetti2}
%
\begin{bbook}[mr]
\bauthor{\bparticle{de} \bsnm{Finetti},~\bfnm{Bruno}\binits{B.}}
(\byear{1990}).
\btitle{Theory of Probability: A Critical Introductory Treatment.
{V}ol. 2}.
\bseries{Wiley Classics Library}.
\blocation{Chichester}: \bpublisher{Wiley}.
\bid{mr={1093667}}
\bptok{imsref}%
\end{bbook}
%
\endbibitem

\bibitem{DelbaenSchachermayer94}
%
\begin{barticle}[mr]
\bauthor{\bsnm{Delbaen},~\bfnm{Freddy}\binits{F.}} \AND
\bauthor{\bsnm{Schachermayer},~\bfnm{Walter}\binits{W.}}
(\byear{1994}).
\btitle{A general version of the fundamental theorem of asset pricing}.
\bjournal{Math. Ann.}
\bvolume{300}
\bpages{463--520}.
\bid{doi={10.1007/BF01450498}, issn={0025-5831}, mr={1304434}}
\bptok{imsref}%
\end{barticle}
%
\endbibitem

\bibitem{DelbaenSchachermayer98}
%
\begin{barticle}[mr]
\bauthor{\bsnm{Delbaen},~\bfnm{F.}\binits{F.}} \AND
\bauthor{\bsnm{Schachermayer},~\bfnm{W.}\binits{W.}}
(\byear{1998}).
\btitle{The fundamental theorem of asset pricing for unbounded stochastic
processes}.
\bjournal{Math. Ann.}
\bvolume{312}
\bpages{215--250}.
\bid{doi={10.1007/s002080050220}, issn={0025-5831}, mr={1671792}}
\bptok{imsref}%
\end{barticle}
%
\endbibitem

\bibitem{DelbaenSchachermayer06}
%
\begin{bbook}[mr]
\bauthor{\bsnm{Delbaen},~\bfnm{Freddy}\binits{F.}} \AND
\bauthor{\bsnm{Schachermayer},~\bfnm{Walter}\binits{W.}}
(\byear{2006}).
\btitle{The Mathematics of Arbitrage}.
\bseries{Springer Finance}.
\blocation{Berlin}: \bpublisher{Springer}.
\bid{mr={2200584}}
\bptok{imsref}%
\end{bbook}
%
\endbibitem

\bibitem{Dellacherie}
%
\begin{bincollection}[mr]
\bauthor{\bsnm{Dellacherie},~\bfnm{C.}\binits{C.}}
(\byear{1971}).
\btitle{Quelques commentaires sur les prolongements de capacit\'es}.
In \bbooktitle{S\'eminaire de {P}robabilit\'es, {V} ({U}niv. {S}trasbourg,
Ann\'ee Universitaire 1969--1970)}.
\bseries{Lecture Notes in Math.}
\bvolume{191}
\bpages{77--81}.
\blocation{Berlin}: \bpublisher{Springer}.
\bid{mr={0382686}}
\bptok{imsref}%
\end{bincollection}
%
\endbibitem

\bibitem{DeprezGerber}
%
\begin{barticle}[mr]
\bauthor{\bsnm{Deprez},~\bfnm{Olivier}\binits{O.}} \AND
\bauthor{\bsnm{Gerber},~\bfnm{Hans~U.}\binits{H.U.}}
(\byear{1985}).
\btitle{On convex principles of premium calculation}.
\bjournal{Insurance Math. Econom.}
\bvolume{4}
\bpages{179--189}.
\bid{doi={10.1016/0167-6687(85)90014-9}, issn={0167-6687}, mr={0797503}}
\bptok{imsref}%
\end{barticle}
%
\endbibitem

\bibitem{Doeblin}
%
\begin{bbook}[mr]
\bauthor{\bsnm{Doeblin},~\bfnm{Wolfgang}\binits{W.}}
(\byear{2000}).
\btitle{Sur L'\'equation de {K}olmogoroff, Par {W}. {D}oeblin}.
\blocation{Paris}: \bpublisher{\'Editions Elsevier}.
\bnote{C. R. Acad. Sci. Paris S{\'e}r. I Math. \textbf{3} 31 (2000), Special
Issue}.
\bptok{imsref}%
\end{bbook}
%
\endbibitem

\bibitem{Dupire1993}
%
\begin{barticle}[author]
\bauthor{\bsnm{Dupire},~\bfnm{Bruno}\binits{B.}}
(\byear{1993}).
\btitle{Model art}.
\bjournal{Risk}
\bvolume{6}
\bpages{118--124}.
\bptok{imsref}%
\end{barticle}
%
\endbibitem

\bibitem{Dupire}
%
\begin{bmisc}[author]
\bauthor{\bsnm{Dupire},~\bfnm{Bruno}\binits{B.}}
(\byear{2009}).
\bhowpublished{Functional {I}t{\^o} Calculus. Bloomberg Portfolio Research
paper.}
\bptok{imsref}%
\end{bmisc}
%
\endbibitem

\bibitem{DynkinParabolic}
%
\begin{barticle}[mr]
\bauthor{\bsnm{Dynkin},~\bfnm{E.~B.}\binits{E.B.}}
(\byear{1992}).
\btitle{Superdiffusions and parabolic nonlinear differential equations}.
\bjournal{Ann. Probab.}
\bvolume{20}
\bpages{942--962}.
\bid{issn={0091-1798}, mr={1159580}}
\bptok{imsref}%
\end{barticle}
%
\endbibitem

\bibitem{RobustBS}
%
\begin{barticle}[mr]
\bauthor{\bsnm{El~Karoui},~\bfnm{Nicole}\binits{N.}},
\bauthor{\bsnm{Jeanblanc-Picqu{\'e}},~\bfnm{Monique}\binits{M.}} \AND
\bauthor{\bsnm{Shreve},~\bfnm{Steven~E.}\binits{S.E.}}
(\byear{1998}).
\btitle{Robustness of the {B}lack and {S}choles formula}.
\bjournal{Math. Finance}
\bvolume{8}
\bpages{93--126}.
\bid{doi={10.1111/1467-9965.00047}, issn={0960-1627}, mr={1609962}}
\bptok{imsref}%
\end{barticle}
%
\endbibitem

\bibitem{ElKarouiQuenez}
%
\begin{barticle}[mr]
\bauthor{\bsnm{El~Karoui},~\bfnm{Nicole}\binits{N.}} \AND
\bauthor{\bsnm{Quenez},~\bfnm{Marie-Claire}\binits{M.C.}}
(\byear{1995}).
\btitle{Dynamic programming and pricing of contingent claims in an incomplete
market}.
\bjournal{SIAM J. Control Optim.}
\bvolume{33}
\bpages{29--66}.
\bid{doi={10.1137/S0363012992232579}, issn={0363-0129}, mr={1311659}}
\bptok{imsref}%
\end{barticle}
%
\endbibitem

\bibitem{FoellmerIto}
%
\begin{bincollection}[author]
\bauthor{\bsnm{F{\"o}llmer},~\bfnm{H.}\binits{H.}}
(\byear{1981}).
\btitle{Calcul d'{I}t\^o sans probabilit\'es}.
In \bbooktitle{Seminar on {P}robability, {XV} ({U}niv. {S}trasbourg,
{S}trasbourg, 1979/1980) ({F}rench)}.
\bseries{Lecture Notes in Math.}
\bvolume{850}
\bpages{143--150}.
\blocation{Berlin}: \bpublisher{Springer}.
\bid{mr={0622559}}
\bptok{imsref}%
\end{bincollection}
%
\endbibitem

\bibitem{FoellmerECM}
%
\begin{bincollection}[mr]
\bauthor{\bsnm{F{\"o}llmer},~\bfnm{Hans}\binits{H.}}
(\byear{2001}).
\btitle{Probabilistic aspects of financial risk}.
In \bbooktitle{European {C}ongress of {M}athematics, {V}ol. {I} ({B}arcelona,
2000)}.
\bseries{Progr. Math.}
\bvolume{201}
\bpages{21--36}.
\blocation{Basel}: \bpublisher{Birkh\"auser}.
\bid{mr={1905311}}
\bptok{imsref}%
\end{bincollection}
%
\endbibitem


\bibitem{F2006}
%
\begin{barticle}[mr]
\bauthor{\bsnm{F{\"o}llmer},~\bfnm{Hans}\binits{H.}} \AND
\bauthor{\bsnm{Gundel},~\bfnm{Anne}\binits{A.}}
(\byear{2006}).
\btitle{Robust projections in the class of martingale measures}.
\bjournal{Illinois J. Math.}
\bvolume{50}
\bpages{439--472 (electronic)}.
\bid{mr={2247836}}
\bptok{imsref}%
\end{barticle}
%
\endbibitem


\bibitem{FoellmerKirman}
%
\begin{barticle}[mr]
\bauthor{\bsnm{F{\"o}llmer},~\bfnm{Hans}\binits{H.}},
\bauthor{\bsnm{Horst},~\bfnm{Ulrich}\binits{U.}} \AND
\bauthor{\bsnm{Kirman},~\bfnm{Alan}\binits{A.}}
(\byear{2005}).
\btitle{Equilibria in financial markets with heterogeneous agents: A
probabilistic perspective}.
\bjournal{J. Math. Econom.}
\bvolume{41}
\bpages{123--155}.
\bid{doi={10.1016/j.jmateco.2004.08.001}, issn={0304-4068}, mr={2120985}}
\bptok{imsref}%
\end{barticle}
%
\endbibitem

\bibitem{FoellmerKabanov}
%
\begin{barticle}[mr]
\bauthor{\bsnm{F{\"o}llmer},~\bfnm{H.}\binits{H.}} \AND
\bauthor{\bsnm{Kabanov},~\bfnm{Yu.~M.}\binits{Y.M.}}
(\byear{1998}).
\btitle{Optional decomposition and {L}agrange multipliers}.
\bjournal{Finance Stoch.}
\bvolume{2}
\bpages{69--81}.
\bid{doi={10.1007/s007800050033}, issn={0949-2984}, mr={1804665}}
\bptok{imsref}%
\end{barticle}
%
\endbibitem

\bibitem{FoellmerLeukert}
%
\begin{barticle}[mr]
\bauthor{\bsnm{F{\"o}llmer},~\bfnm{Hans}\binits{H.}} \AND
\bauthor{\bsnm{Leukert},~\bfnm{Peter}\binits{P.}}
(\byear{2000}).
\btitle{Efficient hedging: Cost versus shortfall risk}.
\bjournal{Finance Stoch.}
\bvolume{4}
\bpages{117--146}.
\bid{doi={10.1007/s007800050008}, issn={0949-2984}, mr={1780323}}
\bptok{imsref}%
\end{barticle}
%
\endbibitem

\bibitem{FoellmerSchiedConv}
%
\begin{barticle}[mr]
\bauthor{\bsnm{F{\"o}llmer},~\bfnm{Hans}\binits{H.}} \AND
\bauthor{\bsnm{Schied},~\bfnm{Alexander}\binits{A.}}
(\byear{2002}).
\btitle{Convex measures of risk and trading constraints}.
\bjournal{Finance Stoch.}
\bvolume{6}
\bpages{429--447}.
\bid{doi={10.1007/s007800200072}, issn={0949-2984}, mr={1932379}}
\bptok{imsref}%
\end{barticle}
%
\endbibitem

\bibitem{FoellmerSchied}
%
\begin{bbook}[mr]
\bauthor{\bsnm{F{\"o}llmer},~\bfnm{Hans}\binits{H.}} \AND
\bauthor{\bsnm{Schied},~\bfnm{Alexander}\binits{A.}}
(\byear{2011}).
\btitle{Stochastic Finance: An Introduction in Discrete Time},
\bedition{third revised and extended} ed.
\blocation{Berlin}: \bpublisher{de Gruyter}.
\bid{mr={2779313}}
\bptok{imsref}%
\end{bbook}
%
\endbibitem

\bibitem{FoellmerSchiedWeber}
%
\begin{bincollection}[author]
\bauthor{\bsnm{F{\"o}llmer},~\bfnm{Hans}\binits{H.}},
\bauthor{\bsnm{Schied},~\bfnm{Alexander}\binits{A.}} \AND
\bauthor{\bsnm{Weber},~\bfnm{Stefan}\binits{S.}}
(\byear{2009}).
\btitle{Robust preferences and robust portfolio choice}.
In \bbooktitle{Mathematical Modelling and Numerical Methods in Finance}
(\beditor{\bfnm{P.}\binits{P.}~\bsnm{Ciarlet}},
\beditor{\bfnm{A.}\binits{A.}~\bsnm{Bensoussan}} \AND
\beditor{\bfnm{Q.}\binits{Q.}~\bsnm{Zhang}}, eds.)
\bvolume{15}
\bpages{29--88}.
\bpublisher{Amsterdam: Elsevier/North-Holland}.
\bptok{imsref}%
\end{bincollection}
%
\endbibitem

\bibitem{FoellmerSchweizer0}
%
\begin{bincollection}[mr]
\bauthor{\bsnm{F{\"o}llmer},~\bfnm{Hans}\binits{H.}} \AND
\bauthor{\bsnm{Schweizer},~\bfnm{Martin}\binits{M.}}
(\byear{1991}).
\btitle{Hedging of contingent claims under incomplete information}.
In \bbooktitle{Applied Stochastic Analysis ({L}ondon, 1989)}.
\bseries{Stochastics Monogr.}
\bvolume{5}
\bpages{389--414}.
\blocation{New York}: \bpublisher{Gordon and Breach}.
\bid{mr={1108430}}
\bptok{imsref}%
\end{bincollection}
%
\endbibitem

\bibitem{FoellmerSchweizer}
%
\begin{barticle}[author]
\bauthor{\bsnm{F{\"o}llmer},~\bfnm{H.}\binits{H.}} \AND
\bauthor{\bsnm{Schweizer},~\bfnm{M.}\binits{M.}}
(\byear{1993}).
\btitle{A microeconomic approach to diffusion models for stock prices}.
\bjournal{Math. Finance}
\bvolume{3}
\bpages{1--23}.
\bptok{imsref}%
\end{barticle}
%
\endbibitem

\bibitem{FoellmerSchweizerEncyclopedia}
%
\begin{bincollection}[author]
\bauthor{\bsnm{F{\"o}llmer},~\bfnm{Hans}\binits{H.}} \AND
\bauthor{\bsnm{Schweizer},~\bfnm{Martin}\binits{M.}}
(\byear{2010}).
\btitle{The minimal martingale measure}.
In \bbooktitle{Encyclopedia of Quantitative Finance}
(\beditor{\bfnm{Rama}\binits{R.}~\bsnm{Cont}}, ed.)
\bpages{1200--1204}.
\blocation{Hoboken, NJ}: \bpublisher{Wiley}.
\bptok{imsref}%
\end{bincollection}
%
\endbibitem

\bibitem{FoellmerSondermann}
%
\begin{bincollection}[mr]
\bauthor{\bsnm{F{\"o}llmer},~\bfnm{Hans}\binits{H.}} \AND
\bauthor{\bsnm{Sondermann},~\bfnm{Dieter}\binits{D.}}
(\byear{1986}).
\btitle{Hedging of nonredundant contingent claims}.
In \bbooktitle{Contributions to Mathematical Economics}
\bpages{205--223}.
\blocation{Amsterdam}: \bpublisher{North-Holland}.
\bid{mr={0902885}}
\bptok{imsref}%
\end{bincollection}
%
\endbibitem

\bibitem{FrittelliRosazzaGianin}
%
\begin{barticle}[author]
\bauthor{\bsnm{Frittelli},~\bfnm{Marco}\binits{M.}} \AND
\bauthor{\bsnm{Rosazza~Gianin},~\bfnm{Emanuela}\binits{E.}}
(\byear{2002}).
\btitle{Putting order in risk measures}.
\bjournal{Journal of Banking \& Finance}
\bvolume{26}
\bpages{1473--1486}.
\bptok{imsref}%
\end{barticle}
%
\endbibitem

\bibitem{FrizVictoir}
%
\begin{bbook}[mr]
\bauthor{\bsnm{Friz},~\bfnm{Peter~K.}\binits{P.K.}} \AND
\bauthor{\bsnm{Victoir},~\bfnm{Nicolas~B.}\binits{N.B.}}
(\byear{2010}).
\btitle{Multidimensional Stochastic Processes as Rough Paths: Theory and
Applications}.
\bseries{Cambridge Studies in Advanced Mathematics}
\bvolume{120}.
\blocation{Cambridge}: \bpublisher{Cambridge Univ. Press}.
\bid{mr={2604669}}
\bptok{imsref}%
\end{bbook}
%
\endbibitem

\bibitem{Gatheralbook}
%
\begin{bbook}[author]
\bauthor{\bsnm{Gatheral},~\bfnm{Jim}\binits{J.}}
(\byear{2006}).
\btitle{The Volatility Surface. A Practitioner's Guide}.
\blocation{Hoboken, NJ}: \bpublisher{Wiley Finance}.
\bptok{imsref}%
\end{bbook}
%
\endbibitem

\bibitem{Gatheral}
%
\begin{barticle}[mr]
\bauthor{\bsnm{Gatheral},~\bfnm{Jim}\binits{J.}}
(\byear{2010}).
\btitle{No-dynamic-arbitrage and market impact}.
\bjournal{Quant. Finance}
\bvolume{10}
\bpages{749--759}.
\bid{doi={10.1080/14697680903373692}, issn={1469-7688}, mr={2741947}}
\bptok{imsref}%
\end{barticle}
%
\endbibitem

\bibitem{GatheralSchiedSurvey}
%
\begin{bmisc}[author]
\bauthor{\bsnm{Gatheral},~\bfnm{Jim}\binits{J.}} \AND
\bauthor{\bsnm{Schied},~\bfnm{Alexander}\binits{A.}}
(\byear{2013}).
\bhowpublished{Dynamical models of market impact and algorithms for order
execution. In \textit{Handbook on Systemic Risk} (J.-P. Fouque and J. Langsam,
eds.). Cambridge:
Cambridge University Press.}
\bptok{imsref}%
\end{bmisc}
%
\endbibitem

\bibitem{GilboaSchmeidler}
%
\begin{barticle}[mr]
\bauthor{\bsnm{Gilboa},~\bfnm{Itzhak}\binits{I.}} \AND
\bauthor{\bsnm{Schmeidler},~\bfnm{David}\binits{D.}}
(\byear{1989}).
\btitle{Maxmin expected utility with nonunique prior}.
\bjournal{J. Math. Econom.}
\bvolume{18}
\bpages{141--153}.
\bid{doi={10.1016/0304-4068(89)90018-9}, issn={0304-4068}, mr={1000102}}
\bptok{imsref}%
\end{barticle}
%
\endbibitem

\bibitem{Goovaerts}
%
\begin{bbook}[mr]
\bauthor{\bsnm{Goovaerts},~\bfnm{M.~J.}\binits{M.J.}},
\bauthor{\bsnm{De~Vylder},~\bfnm{F.}\binits{F.}} \AND
\bauthor{\bsnm{Haezendonck},~\bfnm{J.}\binits{J.}}
(\byear{1984}).
\btitle{Insurance Premiums: Theory and Applications}.
\blocation{Amsterdam}: \bpublisher{North-Holland}.
\bptok{imsref}%
\end{bbook}
%
\endbibitem

\bibitem{HarrisonKreps}
%
\begin{barticle}[mr]
\bauthor{\bsnm{Harrison},~\bfnm{J.~Michael}\binits{J.M.}} \AND
\bauthor{\bsnm{Kreps},~\bfnm{David~M.}\binits{D.M.}}
(\byear{1979}).
\btitle{Martingales and arbitrage in multiperiod securities markets}.
\bjournal{J. Econom. Theory}
\bvolume{20}
\bpages{381--408}.
\bid{doi={10.1016/0022-0531(79)90043-7}, issn={0022-0531}, mr={0540823}}
\bptok{imsref}%
\end{barticle}
%
\endbibitem

\bibitem{Heath}
%
\begin{bmisc}[author]
\bauthor{\bsnm{Heath},~\bfnm{David}\binits{D.}}
(\byear{2000}).
\bhowpublished{Back to the Future. Plenary Lecture, First World
Congress of the
Bachelier Finance Society, Paris.}
\bptok{imsref}%
\end{bmisc}
%
\endbibitem

\bibitem{Hellwig}
%
\begin{barticle}[author]
\bauthor{\bsnm{Hellwig},~\bfnm{Martin}\binits{M.}}
(\byear{2009}).
\btitle{Systemic risk in the financial sector: An analysis of the
subprime-mortgage financial crisis}.
\bjournal{De Economist}
\bvolume{157}
\bpages{129--207}.
\bptok{imsref}%
\end{barticle}
%
\endbibitem

\bibitem{HernandezSchied}
%
\begin{barticle}[mr]
\bauthor{\bsnm{Hern{\'a}ndez-Hern{\'a}ndez},~\bfnm{Daniel}\binits{D.}}
\AND
\bauthor{\bsnm{Schied},~\bfnm{Alexander}\binits{A.}}
(\byear{2007}).
\btitle{A control approach to robust utility maximization with logarithmic
utility and time-consistent penalties}.
\bjournal{Stochastic Process. Appl.}
\bvolume{117}
\bpages{980--1000}.
\bid{doi={10.1016/j.spa.2006.11.005}, issn={0304-4149}, mr={2340875}}
\bptok{imsref}%
\end{barticle}
%
\endbibitem

\bibitem{Hobson}
%
\begin{barticle}[author]
\bauthor{\bsnm{Hobson},~\bfnm{D.~G.}\binits{D.G.}}
(\byear{1998}).
\btitle{Robust hedging of the lookback option}.
\bjournal{Finance Stoch.}
\bvolume{2}
\bpages{329--347}.
\bptok{imsref}%
\end{barticle}
%
\endbibitem

\bibitem{Huber}
%
\begin{bbook}[mr]
\bauthor{\bsnm{Huber},~\bfnm{Peter~J.}\binits{P.J.}}
(\byear{1981}).
\btitle{Robust Statistics}.
\blocation{New York}: \bpublisher{Wiley}.
\bid{mr={0606374}}
\bptok{imsref}%
\end{bbook}
%
\endbibitem

\bibitem{HuberStrassen}
%
\begin{barticle}[mr]
\bauthor{\bsnm{Huber},~\bfnm{Peter~J.}\binits{P.J.}} \AND
\bauthor{\bsnm{Strassen},~\bfnm{Volker}\binits{V.}}
(\byear{1973}).
\btitle{Minimax tests and the {N}eyman--{P}earson lemma for capacities}.
\bjournal{Ann. Statist.}
\bvolume{1}
\bpages{251--263}.
\bid{issn={0090-5364}, mr={0356306}}
\bptok{imsref}%
\end{barticle}
%
\endbibitem

\bibitem{HubermanStanzl}
%
\begin{barticle}[mr]
\bauthor{\bsnm{Huberman},~\bfnm{Gur}\binits{G.}} \AND
\bauthor{\bsnm{Stanzl},~\bfnm{Werner}\binits{W.}}
(\byear{2004}).
\btitle{Price manipulation and quasi-arbitrage}.
\bjournal{Econometrica}
\bvolume{72}
\bpages{1247--1275}.
\bid{doi={10.1111/j.1468-0262.2004.00531.x}, issn={0012-9682}, mr={2064713}}
\bptok{imsref}%
\end{barticle}
%
\endbibitem

\bibitem{JarrowProtter2}
%
\begin{bincollection}[mr]
\bauthor{\bsnm{Jarrow},~\bfnm{Robert~A.}\binits{R.A.}},
\bauthor{\bsnm{Protter},~\bfnm{Philip}\binits{P.}} \AND
\bauthor{\bsnm{Shimbo},~\bfnm{Kazuhiro}\binits{K.}}
(\byear{2007}).
\btitle{Asset price bubbles in complete markets}.
In \bbooktitle{Advances in Mathematical Finance}.
\bseries{Appl. Numer. Harmon. Anal.}
\bpages{97--121}.
\blocation{Boston, MA}: \bpublisher{Birkh\"auser}.
\bid{doi={10.1007/978-0-8176-4545-8_7}, mr={2359365}}
\bptok{imsref}%
\end{bincollection}
%
\endbibitem

\bibitem{JarrowProtter1}
%
\begin{barticle}[mr]
\bauthor{\bsnm{Jarrow},~\bfnm{Robert~A.}\binits{R.A.}},
\bauthor{\bsnm{Protter},~\bfnm{Philip}\binits{P.}} \AND
\bauthor{\bsnm{Shimbo},~\bfnm{Kazuhiro}\binits{K.}}
(\byear{2010}).
\btitle{Asset price bubbles in incomplete markets}.
\bjournal{Math. Finance}
\bvolume{20}
\bpages{145--185}.
\bid{doi={10.1111/j.1467-9965.2010.00394.x}, issn={0960-1627}, mr={2650245}}
\bptok{imsref}%
\end{barticle}
%
\endbibitem

\bibitem{Kabanov}
%
\begin{bincollection}[author]
\bauthor{\bsnm{Kabanov},~\bfnm{Yu.~M.}\binits{Y.M.}}
(\byear{1997}).
\btitle{On the {FTAP} of {K}reps--{D}elbaen--{S}chachermayer}.
In \bbooktitle{Statistics and Control of Stochastic Processes ({M}oscow,
1995/1996)}
\bpages{191--203}.
\blocation{River Edge, NJ}: \bpublisher{World Scientific}.
\bptok{imsref}%
\end{bincollection}
%
\endbibitem

\bibitem{KaratzasLehoczkyShreve}
%
\begin{barticle}[mr]
\bauthor{\bsnm{Karatzas},~\bfnm{Ioannis}\binits{I.}},
\bauthor{\bsnm{Lehoczky},~\bfnm{John~P.}\binits{J.P.}} \AND
\bauthor{\bsnm{Shreve},~\bfnm{Steven~E.}\binits{S.E.}}
(\byear{1987}).
\btitle{Optimal portfolio and consumption decisions for a ``small
investor'' on
a finite horizon}.
\bjournal{SIAM J. Control Optim.}
\bvolume{25}
\bpages{1557--1586}.
\bid{doi={10.1137/0325086}, issn={0363-0129}, mr={0912456}}
\bptok{imsref}%
\end{barticle}
%
\endbibitem

\bibitem{KaratzasShreveMF}
%
\begin{bbook}[mr]
\bauthor{\bsnm{Karatzas},~\bfnm{Ioannis}\binits{I.}} \AND
\bauthor{\bsnm{Shreve},~\bfnm{Steven~E.}\binits{S.E.}}
(\byear{1998}).
\btitle{Methods of Mathematical Finance}.
\bseries{Applications of Mathematics (New York)}
\bvolume{39}.
\blocation{New York}: \bpublisher{Springer}.
\bid{mr={1640352}}
\bptok{imsref}%
\end{bbook}
%
\endbibitem

\bibitem{Kirman}
%
\begin{barticle}[author]
\bauthor{\bsnm{Kirman},~\bfnm{Alan}\binits{A.}}
(\byear{2010}).
\btitle{The economic crisis is a crisis for economic theory}.
\bjournal{CESifo Economic Studies}
\bvolume{56}
\bpages{498--535}.
\bptok{imsref}%
\end{barticle}
%
\endbibitem

\bibitem{KSS}
%
\begin{bmisc}[author]
\bauthor{\bsnm{Kl{\"o}ck},~\bfnm{Florian}\binits{F.}},
\bauthor{\bsnm{Schied},~\bfnm{Alexander}\binits{A.}} \AND
\bauthor{\bsnm{Sun},~\bfnm{Yuemeng}\binits{Y.}}
(\byear{2011}).
\bhowpublished{Price manipulation in a market impact model with dark pool.
Unpublished manuscript.}
\bptok{imsref}%
\end{bmisc}
%
\endbibitem

\bibitem{Knight}
%
\begin{bbook}[author]
\bauthor{\bsnm{Knight},~\bfnm{F.}\binits{F.}}
(\byear{1921}).
\btitle{Risk, Uncertainty, and Profit}.
\blocation{Boston}: \bpublisher{Houghton Mifflin}.
\bptok{imsref}%
\end{bbook}
%
\endbibitem

\bibitem{KramkovSchachermayer1}
%
\begin{barticle}[mr]
\bauthor{\bsnm{Kramkov},~\bfnm{D.}\binits{D.}} \AND
\bauthor{\bsnm{Schachermayer},~\bfnm{W.}\binits{W.}}
(\byear{1999}).
\btitle{The asymptotic elasticity of utility functions and optimal investment
in incomplete markets}.
\bjournal{Ann. Appl. Probab.}
\bvolume{9}
\bpages{904--950}.
\bid{doi={10.1214/aoap/1029962818}, issn={1050-5164}, mr={1722287}}
\bptok{imsref}%
\end{barticle}
%
\endbibitem

\bibitem{KramkovSchachermayer2}
%
\begin{barticle}[mr]
\bauthor{\bsnm{Kramkov},~\bfnm{D.}\binits{D.}} \AND
\bauthor{\bsnm{Schachermayer},~\bfnm{W.}\binits{W.}}
(\byear{2003}).
\btitle{Necessary and sufficient conditions in the problem of optimal
investment in incomplete markets}.
\bjournal{Ann. Appl. Probab.}
\bvolume{13}
\bpages{1504--1516}.
\bid{doi={10.1214/aoap/1069786508}, issn={1050-5164}, mr={2023886}}
\bptok{imsref}%
\end{barticle}
%
\endbibitem

\bibitem{Kramkov}
%
\begin{barticle}[mr]
\bauthor{\bsnm{Kramkov},~\bfnm{D.~O.}\binits{D.O.}}
(\byear{1996}).
\btitle{Optional decomposition of supermartingales and hedging contingent
claims in incomplete security markets}.
\bjournal{Probab. Theory Related Fields}
\bvolume{105}
\bpages{459--479}.
\bid{doi={10.1007/BF01191909}, issn={0178-8051}, mr={1402653}}
\bptok{imsref}%
\end{barticle}
%
\endbibitem

\bibitem{KratzSchoeneborn}
%
\begin{bmisc}[author]
\bauthor{\bsnm{Kratz},~\bfnm{Peter}\binits{P.}} \AND
\bauthor{\bsnm{Sch{\"{o}}neborn},~\bfnm{Torsten}\binits{T.}}
(\byear{2010}).
\bhowpublished{Optimal liquidation in dark pools. Unpublished manuscript.}
\bptok{imsref}%
\end{bmisc}
%
\endbibitem

\bibitem{Kreps}
%
\begin{bbook}[author]
\bauthor{\bsnm{Kreps},~\bfnm{David~M.}\binits{D.M.}}
(\byear{1979}).
\btitle{Three Essays on Capital Markets}.
\blocation{Stanford University}:  \bpublisher{Institute for Mathematical Studies in the Social
Sciences}.
\bnote{Reprinted in \textit{Revista Espa\~{n}ola de Economica} \textbf{4} (1987), 111--146.}
\bptok{imsref}%
\end{bbook}
%
\endbibitem

\bibitem{Lehalle}
%
\begin{bmisc}[author]
\bauthor{\bsnm{Lehalle},~\bfnm{Charles-Albert}\binits{C.A.}}
(\byear{2013}).
\bhowpublished{Market microstructure knowledge needed to control an intra-day
trading process. In \textit{Handbook on Systemic Risk} (J.-P. Fouque and J. Langsam,
eds.). Cambridge:
Cambridge University Press.}
\bptok{imsref}%
\end{bmisc}
%
\endbibitem

\bibitem{LyonsQian}
%
\begin{bbook}[mr]
\bauthor{\bsnm{Lyons},~\bfnm{Terry}\binits{T.}} \AND
\bauthor{\bsnm{Qian},~\bfnm{Zhongmin}\binits{Z.}}
(\byear{2002}).
\btitle{System Control and Rough Paths}.
\bseries{Oxford Mathematical Monographs}.
\blocation{Oxford}: \bpublisher{Oxford Univ. Press}.
\bid{doi={10.1093/acprof:oso/9780198506485.001.0001}, mr={2036784}}
\bptok{imsref}%
\end{bbook}
%
\endbibitem

\bibitem{Maccheronietal}
%
\begin{barticle}[mr]
\bauthor{\bsnm{Maccheroni},~\bfnm{Fabio}\binits{F.}},
\bauthor{\bsnm{Marinacci},~\bfnm{Massimo}\binits{M.}} \AND
\bauthor{\bsnm{Rustichini},~\bfnm{Aldo}\binits{A.}}
(\byear{2006}).
\btitle{Ambiguity aversion, robustness, and the variational
representation of
preferences}.
\bjournal{Econometrica}
\bvolume{74}
\bpages{1447--1498}.
\bid{doi={10.1111/j.1468-0262.2006.00716.x}, issn={0012-9682}, mr={2268407}}
\bptok{imsref}%
\end{barticle}
%
\endbibitem

\bibitem{Merton}
%
\begin{barticle}[mr]
\bauthor{\bsnm{Merton},~\bfnm{Robert~C.}\binits{R.C.}}
(\byear{1973}).
\btitle{Theory of rational option pricing}.
\bjournal{Bell J. Econom. and Management Sci.}
\bvolume{4}
\bpages{141--183}.
\bid{issn={0741-6261}, mr={0496534}}
\bptok{imsref}%
\end{barticle}
%
\endbibitem

\bibitem{Mittal}
%
\begin{barticle}[author]
\bauthor{\bsnm{Mittal},~\bfnm{Hitesh}\binits{H.}}
(\byear{2008}).
\btitle{Are you playing in a toxic dark pool? {A} guide to preventing
information leakage}.
\bjournal{Journal of {T}rading}
\bvolume{3}
\bpages{20--33}.
\bptok{imsref}%
\end{barticle}
%
\endbibitem

\bibitem{Monroe1}
%
\begin{barticle}[mr]
\bauthor{\bsnm{Monroe},~\bfnm{Itrel}\binits{I.}}
(\byear{1972}).
\btitle{On embedding right continuous martingales in {B}rownian motion}.
\bjournal{Ann. Math. Statist.}
\bvolume{43}
\bpages{1293--1311}.
\bid{issn={0003-4851}, mr={0343354}}
\bptok{imsref}%
\end{barticle}
%
\endbibitem

\bibitem{Monroe2}
%
\begin{barticle}[mr]
\bauthor{\bsnm{Monroe},~\bfnm{Itrel}\binits{I.}}
(\byear{1978}).
\btitle{Processes that can be embedded in {B}rownian motion}.
\bjournal{Ann. Probab.}
\bvolume{6}
\bpages{42--56}.
\bid{mr={0455113}}
\bptok{imsref}%
\end{barticle}
%
\endbibitem

\bibitem{MusielaZariphopoulou1}
%
\begin{barticle}[mr]
\bauthor{\bsnm{Musiela},~\bfnm{M.}\binits{M.}} \AND
\bauthor{\bsnm{Zariphopoulou},~\bfnm{T.}\binits{T.}}
(\byear{2009}).
\btitle{Portfolio choice under dynamic investment performance criteria}.
\bjournal{Quant. Finance}
\bvolume{9}
\bpages{161--170}.
\bid{doi={10.1080/14697680802624997}, issn={1469-7688}, mr={2512986}}
\bptok{imsref}%
\end{barticle}
%
\endbibitem

\bibitem{MusielaZariphopoulou2}
%
\begin{bincollection}[mr]
\bauthor{\bsnm{Musiela},~\bfnm{Marek}\binits{M.}} \AND
\bauthor{\bsnm{Zariphopoulou},~\bfnm{Thaleia}\binits{T.}}
(\byear{2010}).
\btitle{Stochastic partial differential equations and portfolio choice}.
In \bbooktitle{Contemporary Quantitative Finance}
\bpages{195--216}.
\blocation{Berlin}: \bpublisher{Springer}.
\bid{doi={10.1007/978-3-642-03479-4_11}, mr={2732847}}
\bptok{imsref}%
\end{bincollection}
%
\endbibitem

\bibitem{Neuberger}
%
\begin{barticle}[author]
\bauthor{\bsnm{Neuberger},~\bfnm{Anthony}\binits{A.}}
(\byear{1994}).
\btitle{The log contract}.
\bjournal{The Journal of Portfolio Management}
\bvolume{20}
\bpages{74--80}.
\bptok{imsref}%
\end{barticle}
%
\endbibitem

\bibitem{ow}
%
\begin{barticle}[author]
\bauthor{\bsnm{Obizhaeva},~\bfnm{Anna}\binits{A.}} \AND
\bauthor{\bsnm{Wang},~\bfnm{Jiang}\binits{J.}}
(\byear{2013}).
\btitle{Optimal trading strategy and supply/demand dynamics.}
\bjournal{J. Financial Markets}
\bvolume{16}
\bpages{1--32}.
\bptok{imsref}%
\end{barticle}
%
\endbibitem


\bibitem{P1908}
%
\begin{barticle}[auto]
\bauthor{\bsnm{Poincar\'e},~\bfnm{H.}\binits{H.}}
(\byear{1908}).
\btitle{Science et m\'ethode.}
\bjournal{Revue scient. (5)}
\bvolume{10}
\bpages{417--423}.
\bptok{imsref}%
\end{barticle}
%
\endbibitem


\bibitem{RevuzYor}
%
\begin{bbook}[mr]
\bauthor{\bsnm{Revuz},~\bfnm{Daniel}\binits{D.}} \AND
\bauthor{\bsnm{Yor},~\bfnm{Marc}\binits{M.}}
(\byear{1999}).
\btitle{Continuous Martingales and {B}rownian Motion},
\bedition{3rd} ed.
\bseries{Grundlehren der Mathematischen Wissenschaften [Fundamental Principles
of Mathematical Sciences]}
\bvolume{293}.
\blocation{Berlin}: \bpublisher{Springer}.
\bid{mr={1725357}}
\bptok{imsref}%
\end{bbook}
%
\endbibitem

\bibitem{Samuelson}
%
\begin{barticle}[author]
\bauthor{\bsnm{Samuelson},~\bfnm{Paul~A.}\binits{P.A.}}
(\byear{1965}).
\btitle{Proof that properly anticipated prices fluctuate randomly}.
\bjournal{Industrial Management Review}
\bvolume{6}.
\bptok{imsref}%
\end{barticle}
%
\endbibitem

\bibitem{Savage}
%
\begin{bbook}[mr]
\bauthor{\bsnm{Savage},~\bfnm{Leonard~J.}\binits{L.J.}}
(\byear{1972}).
\btitle{The Foundations of Statistics},
\bedition{revised} ed.
\blocation{New York}: \bpublisher{Dover Publications Inc.}
\bid{mr={0348870}}
\bptok{imsref}%
\end{bbook}
%
\endbibitem

\bibitem{Schervish}
%
\begin{barticle}[mr]
\bauthor{\bsnm{Schervish},~\bfnm{Mark~J.}\binits{M.J.}},
\bauthor{\bsnm{Seidenfeld},~\bfnm{Teddy}\binits{T.}} \AND
\bauthor{\bsnm{Kadane},~\bfnm{Joseph~B.}\binits{J.B.}}
(\byear{2008}).
\btitle{The fundamental theorems of prevision and asset pricing}.
\bjournal{Internat. J. Approx. Reason.}
\bvolume{49}
\bpages{148--158}.
\bid{doi={10.1016/j.ijar.2007.06.012}, issn={0888-613X}, mr={2454836}}
\bptok{imsref}%
\end{barticle}
%
\endbibitem

\bibitem{SchiedMOR}
%
\begin{barticle}[mr]
\bauthor{\bsnm{Schied},~\bfnm{Alexander}\binits{A.}}
(\byear{2005}).
\btitle{Optimal investments for robust utility functionals in complete market
models}.
\bjournal{Math. Oper. Res.}
\bvolume{30}
\bpages{750--764}.
\bid{doi={10.1287/moor.1040.0138}, issn={0364-765X}, mr={2161208}}
\bptok{imsref}%
\end{barticle}
%
\endbibitem

\bibitem{SchiedFS}
%
\begin{barticle}[mr]
\bauthor{\bsnm{Schied},~\bfnm{Alexander}\binits{A.}}
(\byear{2007}).
\btitle{Optimal investments for risk- and ambiguity-averse preferences: A~duality approach}.
\bjournal{Finance Stoch.}
\bvolume{11}
\bpages{107--129}.
\bid{doi={10.1007/s00780-006-0024-2}, issn={0949-2984}, mr={2284014}}
\bptok{imsref}%
\end{barticle}
%
\endbibitem

\bibitem{SchiedFuel}
%
\begin{bmisc}[author]
\bauthor{\bsnm{Schied},~\bfnm{Alexander}\binits{A.}}
(\byear{2013}).
\bhowpublished{A control problem with fuel constraint and {D}awson--{W}atanabe
superprocesses. \textit{Ann. Appl. Probab.} To appear.}
\bptok{imsref}%
\end{bmisc}
%
\endbibitem

\bibitem{SchiedStadje}
%
\begin{barticle}[mr]
\bauthor{\bsnm{Schied},~\bfnm{Alexander}\binits{A.}} \AND
\bauthor{\bsnm{Stadje},~\bfnm{Mitja}\binits{M.}}
(\byear{2007}).
\btitle{Robustness of delta hedging for path-dependent options in local
volatility models}.
\bjournal{J. Appl. Probab.}
\bvolume{44}
\bpages{865--879}.
\bid{doi={10.1239/jap/1197908810}, issn={0021-9002}, mr={2382931}}
\bptok{imsref}%
\end{barticle}
%
\endbibitem

\bibitem{Schmeidler}
%
\begin{barticle}[mr]
\bauthor{\bsnm{Schmeidler},~\bfnm{David}\binits{D.}}
(\byear{1986}).
\btitle{Integral representation without additivity}.
\bjournal{Proc. Amer. Math. Soc.}
\bvolume{97}
\bpages{255--261}.
\bid{doi={10.2307/2046508}, issn={0002-9939}, mr={0835875}}
\bptok{imsref}%
\end{barticle}
%
\endbibitem

\bibitem{Schoeneborn}
%
\begin{bmisc}[author]
\bauthor{\bsnm{Sch{\"o}neborn},~\bfnm{Torsten}\binits{T.}}
(\byear{2008}).
\bhowpublished{Trade execution in illiquid markets. {O}ptimal stochastic
control and multi-agent equilibria. Ph.D. thesis, TU Berlin}.
\bptok{imsref}%
\end{bmisc}
%
\endbibitem

\bibitem{SchoenebornSchied}
%
\begin{bmisc}[author]
\bauthor{\bsnm{Sch{\"o}neborn},~\bfnm{Torsten}\binits{T.}} \AND
\bauthor{\bsnm{Schied},~\bfnm{Alexander}\binits{A.}}
(\byear{2009}).
\bhowpublished{Liquidation in the face of adversity: Stealth vs. sunshine
trading. Unpublished manuscript}.
\bptok{imsref}%
\end{bmisc}
%
\endbibitem

\bibitem{Schweizer}
%
\begin{bincollection}[author]
\bauthor{\bsnm{Schweizer},~\bfnm{Martin}\binits{M.}}
(\byear{2010}).
\btitle{Mean-variance hedging}.
In \bbooktitle{Encyclopedia of Quantitative Finance}
(\beditor{\bfnm{Rama}\binits{R.}~\bsnm{Cont}}, ed.)
\bpages{1177--1181}.
\bpublisher{Wiley}.
\bptok{imsref}%
\end{bincollection}
%
\endbibitem

\bibitem{Sondermann}
%
\begin{bbook}[mr]
\bauthor{\bsnm{Sondermann},~\bfnm{Dieter}\binits{D.}}
(\byear{2006}).
\btitle{Introduction to Stochastic Calculus for Finance: A New Didactic
Approach}.
\bseries{Lecture Notes in Economics and Mathematical Systems}
\bvolume{579}.
\blocation{Berlin}: \bpublisher{Springer}.
\bid{mr={2254170}}
\bptok{imsref}%
\end{bbook}
%
\endbibitem

\bibitem{StoikovZariphopoulou}
%
\begin{barticle}[author]
\bauthor{\bsnm{Stoikov},~\bfnm{Sasha~F.}\binits{S.F.}} \AND
\bauthor{\bsnm{Zariphopoulou},~\bfnm{Thaleia}\binits{T.}}
(\byear{2005}).
\btitle{Dynamic asset allocation and consumption choice in incomplete markets}.
\bjournal{Australian Economic Papers}
\bvolume{44}
\bpages{414--454}.
\bptok{imsref}%
\end{barticle}
%
\endbibitem

\bibitem{Turner}
%
\begin{bmisc}[author]
\bauthor{\bsnm{Turner},~\bfnm{Adair}\binits{A.}}
(\byear{2009}).
\bhowpublished{The {T}urner Review: A regulatory response to the global banking
crisis. FSA, March}.
\bptok{imsref}%
\end{bmisc}
%
\endbibitem

\bibitem{NeumannMorgenstern}
%
\begin{bbook}[mr]
\bauthor{\bparticle{von} \bsnm{Neumann},~\bfnm{John}\binits{J.}} \AND
\bauthor{\bsnm{Morgenstern},~\bfnm{Oskar}\binits{O.}}
(\byear{1980}).
\btitle{Theory of Games and Economic Behavior},
\bedition{3rd} ed.
\blocation{Princeton, NJ}: \bpublisher{Princeton Univ. Press}.
\bid{mr={0565457}}
\bptok{imsref}%
\end{bbook}
%
\endbibitem

\bibitem{WeberRosenow}
%
\begin{barticle}[author]
\bauthor{\bsnm{Weber},~\bfnm{P.}\binits{P.}} \AND
\bauthor{\bsnm{Rosenow},~\bfnm{B.}\binits{B.}}
(\byear{2005}).
\btitle{Order book approach to price impact}.
\bjournal{Quant. Finance}
\bvolume{5}
\bpages{357--364}.
\bptok{imsref}%
\end{barticle}
%
\endbibitem

\bibitem{Yan}
%
\begin{binproceedings}[mr]
\bauthor{\bsnm{Yan},~\bfnm{Jia-An}\binits{J.A.}}
(\byear{2002}).
\btitle{A numeraire-free and original probability based framework for financial
markets}.
In \bbooktitle{Proceedings of the {I}nternational {C}ongress of
{M}athematicians, {V}ol. {III} ({B}eijing, 2002)}
\bpages{861--871}.
\blocation{Beijing}: \bpublisher{Higher Ed. Press}.
\bid{mr={1957586}}
\bptok{imsref}%
\end{binproceedings}
%
\endbibitem

\end{thebibliography}
\end{document}